\documentclass[aps,floatfix,epsfig]{revtex4}
\usepackage{pdfpages}
\usepackage{epsf}
\usepackage{amsmath}
\usepackage{eqnarray,amsmath}
\usepackage{epstopdf}
\usepackage{mathrsfs}
\usepackage{natbib}
\usepackage{amssymb,latexsym}
\usepackage{dcolumn}
\usepackage{graphicx}
\newcommand{\scri}{\mathscr{I}}
\def\be{\begin{equation}}
\def\ee{\end{equation}}
\def\ba{\begin{eqnarray}}
\def\ea{\end{eqnarray}}

\newcommand{\vac}{|\text{vac}\rangle}

\newcommand{\bvac}{|\overline{\text{vac}}\rangle}
\newcommand{\bvacbra}{\langle\overline{\text{vac}}|}
\newcommand{\bu}{\bar{u}}
\newcommand{\bv}{\bar{v}}
\newcommand{\bm}{\bar{\mu}}
\newcommand{\bn}{\bar{\nu}}

\newcommand{\tmn}[1]{\langle T_{#1} \rangle}

\newcommand{\ttmn}[1]{\langle \widetilde{T}_{#1} \rangle}
\newcommand{\tg}{\widetilde{g}}
\newcommand{\tW}{\widetilde{W}}

\newcommand{\defeq}{:=}

\begin{document}
\title{Particle creation from the quantum stress tensor}

 \author{Javad T. Firouzjaee${}^{1,2}$}
\affiliation{${}^1$School of Astronomy, Institute for Research in Fundamental Sciences (IPM), P. O. Box 19395-5531, Tehran, Iran }
 \email{j.taghizadeh.f@ipm.ir}
\author{George F R Ellis${}^{2}$}
\affiliation{${}^2$Mathematics and Applied Mathematics Department, University of Cape Town, Rondebosch, Cape Town 7701, South Africa}
\email{gfrellisf@gmail.com}
\begin{abstract}

\textbf{Abstract:} Among the different methods to derive particle creation, finding the quantum stress tensor expectation value 
gives a covariant quantity which can be used for examining the back-reaction issue. However this tensor also includes vacuum polarization in a way that depends on the vacuum chosen. Here we review different aspects of particle creation by looking at energy conservation and at the quantum stress tensor. It will be shown that in the case of general spherically symmetric black holes that have a \emph{dynamical horizon}, as  occurs in a cosmological context, one cannot have pair creation on the horizon because this violates energy conservation. This confirms the results obtained in other ways in a previous paper \cite{javad-ellis}. Looking at the expectation value of the quantum stress tensor with three different definitions of the vacuum state, we study the nature of particle creation and vacuum polarization in  black hole and cosmological models, and the associated stress energy tensors. We show that the thermal temperature that is calculated from the particle flux given by the quantum stress tensor is compatible with the temperature determined by the affine null parameter approach. Finally, it will be shown that in the spherically symmetric dynamic case, we can neglect the backscattering term and only consider the s-waves term near the future apparent horizon.

\end{abstract}


\maketitle

\tableofcontents

\vspace{0.25in}

\section{Introduction}
Studying Hawking radiation for cosmological black holes,  which are embedded in a cosmological background, is not trivial \cite{ellis-13}. One of the main problems is that due to the dynamical nature of the infalling 
matter in this case, we have to consider dynamical black hole horizons in the real expanding universe \cite{ellis-ritu-mansouri}, which have different properties  than stationary event horizons.\\

Although about 40 years have passed since  Hawking's paper showed that Schwarzschild black holes emit thermal radiation, some aspects of this process are not understood totally. Hawking used quantum field theory in a curved spacetime to calculate the expectation value of the particle number operator \cite{hawking-75}. However  the expectation value of this 
operator does not expresses pure particle creation in a curved space time. We will discuss how quantum vacuum polarization may also be included in this quantity. Usually in the case that the spacetime is asymptotically Minkowskian in both the far past and future we  can use the Bogoliubov transformation  relating the future and past vacua to read off the pure particle creation in the spacetime: $\langle 0 | N_k| 0 \rangle =\langle 0 |a_k^{\dag} a_k | 0 \rangle=\Sigma_j |\beta_{jk}|^2  $. However the real universe is not asymptotically flat in either the future or the past.\\

Hawking's quantum field theory approach to black hole radiation, which applies to late time stationary black holes, is not a suitable method for calculating the Hawking temperature in the case of a fully dynamical black hole, where one has to solve the field equations in a changing background. There are alternative approaches allowing one to calculate Hawking radiation in a Schwarzschild spacetime without using the field equations, such  as finding the related vacuum via the tunnelling method \cite{tunneling}. One can extend these methods to calculating quantum fields in the dynamical case  \cite{visser-03, vanzo-09}, giving us the ability to determine Hawking radiation for dynamical black holes in a cosmological context \cite{javad-ellis, javad-radiation}. However differentiating the different quantum effects, like vacuum polarization and  particle creation, needs careful study of the expectation value for the quantum stress tensor $\langle \psi | T_{\mu \nu} | \psi \rangle  $.\\

It is known that the particle concept in quantum field theory is a global concept. The particle modes are defined on the whole spacetime, so that particular observers specify them by a field mode decomposition. There is also the number operator describing the response of a particle detector,   which depends on an observer's past history. To get a more usable definition of the particle state, one needs to construct locally-defined quantities.  The best candidate for studying particle creation and other quantum effects locally is the expectation stress tensor value $\langle \psi | T_{\mu \nu} | \psi \rangle  $ \cite{davies-76}, which assumes a particular value at each point $x$ of spacetime. This stress-tensor is objective in the sense that for a fixed state $ | \psi \rangle$, the results of different measuring devices can be related in the familiar fashion by the usual tensor transformation rules. For example if  $\langle \psi | T_{\mu \nu} | \psi \rangle(x)=0 $ for one observer, it will vanish for all observers. However this term does not express pure particle creation, it includes also 
vacuum polarization; and the way it does so depends on the choice of vacuum.\\

There are several effects in quantum field theory one might take into account in a curved spacetime:  zero point energy, the static Casimir effect, the dynamical Casimir effect (which is like a moving mirror), and the Schwinger effect. Particle creation can be attributed to the Hawking effect, the dynamical Casimir effect, 
and the Schwinger effect.  Basically, Hawking particle creation is thermal radiation due to the black hole apparent horizon, which is observed at a large distance (or alternatively is due to an adiabatic condition between affine coordinates on past and future null infinity \cite{visser-10}); and the apparent horizon is the event horizon in the static case, so then the radiation is associated with the event horizon. On the other hand, the dynamical Casimir effect is the production of particles and energy from an accelerated moving mirror, so it is due to dynamical boundary conditions (see \cite{brodag-book} for a comprehensive review). Finally, the 
 Schwinger effect is a non-perturbative QED phenomenon. It is the spontaneous production of
$e^+$ and  $e^-$ pairs in the presence of strong (usually constant) electric fields. \\

To include all particle creation effects in the dynamical case of black hole collapse, we first need to solve the wave equation with suitable boundary condition in the presence of any electromagnetic field; second, find an appropriate vacuum corresponding to the collapsing model (like the Unruh vacuum); and last,  read off the general particle creation from  the number density operator $\langle \psi |  N=a^{\dagger}_k a_k | \psi \rangle$ and expectation stress tensor $ \langle \psi | T_{\mu \nu}| \psi \rangle$. Considering the Schwinger and dynamical Casimir particle creation effects are beyond the scope of this paper. Here we just discuss quantum effects due to the curved  classical spacetime background.\\

The basic point of this work is that it  gives a compact review of different aspects of particle creation, particularly in terms of deriving the quantum stress tensor, in ways which can be applied to non-stationary metrics as well as the stationary case. This is what is needed for backreaction studies in the dynamic black hole case.\\ 

This paper is organized as follow: We will show the relation between energy conservation for the particle creation mechanism and the type of the horizon in section II. In Section III, we discuss the particle creation concept via expectation values  of stress tensor. Section IV considers particle creation concepts in cosmological models. In Section V we discuss different aspects of these  quantum effect for compact stars and black hole models. We then conclude in section VI.

\section{Energy conservation for vacuum creation and annihilation}
As is known, if the total energy change for particle production does not vanish, this process is forbidden by energy
conservation. We consider here necessary conditions so that energy conservation is obeyed for virtual particle pair creation and annihilation. This puts constraints on particle creation in dynamical black hole spacetimes. Although semi-classical physics breaks the energy conservation law up to the fluctuations allowed by Heisenberg’s uncertainty principle $\Delta t \Delta E= \hbar$, for long-lived real particles we cannot neglect  energy conservation violation, so must demand $ \Delta E \ll \frac{\hbar}{\Delta t}$. \\

It can be shown that if there is a Killing horizon, one does not 
have to generate energy for particle pair creation; thus there is no problem in the static case. But in the dynamic context, for example when matter and radiation fall into the black hole, there is no Killing horizon,  
instead there is an apparent horizon that is spacelike when such infall is significant \cite{javad-ellis}. We show here that this  
 makes pair production inconsistent with energy conservation, and so Hawking radiation is then prohibited.  
However at late times it becomes an isolated horizon; 
then, one can again have particle pair creation without violating energy conservation. This confirms the results of \cite{javad-ellis}.\\  

\textbf{The stationary case}:  Consider the possibility of particle creation by a stationary gravitational field. The energy of a particle in such a field is $E=- 
p_\mu \xi^\mu$ where $p^\mu$ is the four-momentum of the particle, and $\xi^\mu$ is the Killing vector field.  The energy $E$ of a particle is always 
positive  outside the black hole horizon, where the  Killing vector is timelike.  
The Killing vector is spacelike inside the Killing 
horizon $\xi^2=0$, and the energy is negative there. Therefore, this allows particle pair creation just around the Killing horizon. On the other hand, we 
know that a Killing horizon in a stationary spacetime is necessarily an event horizon (see Hawking and Ellis \cite{ellis-book}, proposition 9.3.6). 
Hence, one can expect particle creation in a stationary spacetime which contains a black hole \cite{novikov-book}. \\

\textbf{The dynamical case} Now the question is how does energy conservation work for pair creation around a dynamical black hole.
It is not possible to define a conserved quantity $E$ for an evolving space time, as it does not have a Killing vector field. But locally for a 
vector field $U^\mu(x)$ that is timelike 
outside the black hole, 
we can define the particle's energy relative to an observer with 4-velocity $U^\mu$ as $E=- p_\mu U^\mu$. For instance, in a general spherical symmetry spacetime,  one can define the energy $E=- p_\mu K^\mu$ of the particle relative to the Kodama vector  $K^\mu$ \cite{kodama-80}, which in the stationary  case becomes the same as the Killing vector. 

\begin{figure}[h]
\begin{center}
\includegraphics[scale = 0.5]{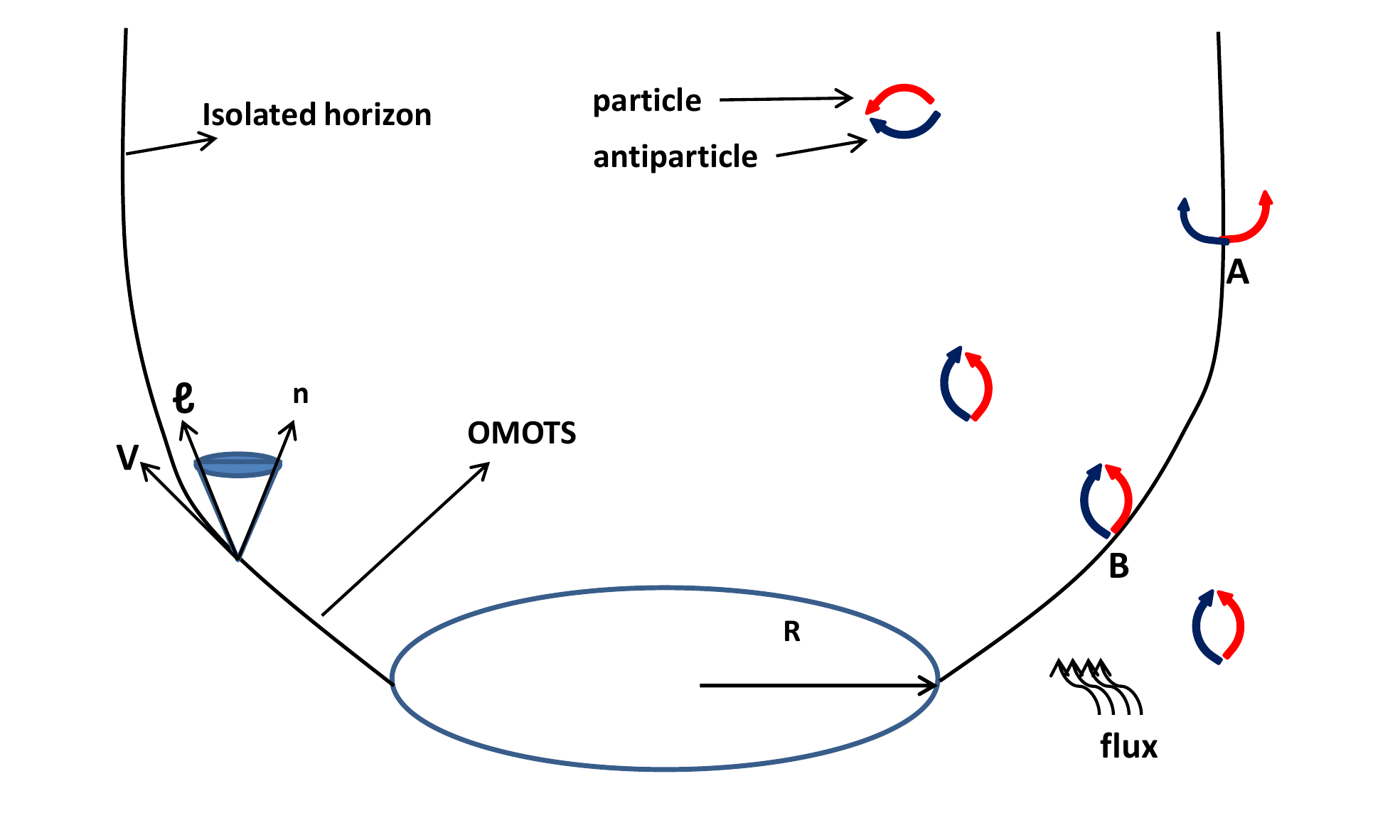}
\hspace*{10mm} \caption{ \label{particle}
 \textit{Particle creation and annihilation around the apparent horizon for the case if an influx that decays away at late times.
 If a pair particle create at point B, they will have the same sign energy and the cannot live long because they violate the energy condition. But when they are created on the isolated horizon at point A the two particle have different sign of energy which keep the energy conservation. Note that we did not include radiation back reaction in this Figure, which is adapted from arXiv:gr-qc/0308033.} }
\end{center}
\end{figure}

Consider pair creation around the OMOTS (Outer Marginally Trapped 3-Surface) when there is infalling radiation or matter, so this surface is spacelike \cite{ellis-ritu-mansouri}. Since the particle -- antiparticle pair will be located inside the horizon (point B in the Fig.(\ref{particle})) and will each have energy defined relative to the same timelike 
vector, the energy $E$ for both particles will have the same sign. As a result, any long-lived particle creation violates the energy conservation around the OMOTS. Hence Hawking radiation will not occur there \cite{javad-ellis}.\\
 
\textbf{ The isolated horizon case 
} When the matter flux becomes very small, the black hole apparent horizon becomes an isolated horizon \cite{ashtekar-98} where its tangent vector becomes a null vector  $\ell^\mu$ and it becomes a null surface. Note that a Killing horizon assumes the existence of a Killing vector field in some neighborhood of the horizon, but the isolated horizon is  defined only in terms of the intrinsic and extrinsic geometry of the horizon itself, where $\ell^\mu$ is a Killing vector for the intrinsic geometry of 
 horizon \cite{ashtekar-98}. To consider energy conservation for pair creation around the isolated horizon, we limit ourselves to the general 
 spherically symmetric dynamic black hole. Consider that the collapsing  fluid within a compact spherically symmetric
spacetime region will be described by the following metric in the comoving coordinates $(t,r,\theta,\varphi)$:
\begin{equation} \label{gltbm}
ds^{2}=-e^{2\nu(t,r)}dt^{2}+e^{2\psi(t,r)}dr^{2}+R(t,r)^{2}d\Omega^{2}.
\end{equation}
 The Kodama vector for this metric is
 \be \label{kodama}
 K^\mu=e^{-(\nu+\psi)}(R', -\dot{R}, 0, 0).
 \ee
Using the Einstein equations one can show that the norm of the Kodama vector is
 \be
 K_\mu K^\mu= (\frac{2M}{R}-1),
 \ee
 where $M$ is the Misner-Sharp mass for the spherically symmetric model \cite{javad-penn}. This vector is spacelike inside the apparent horizon and timelike outside it. 
 Using the energy definition $E=- p_\mu K^\mu$ , one can show that the energy of the particle outside the isolated horizon \cite{ashtekar-98} is positive and the energy of the particle inside is negative (point A in the Fig.(\ref{particle})). Hence, one can have pair particle creation for isolated horizons without violating energy conservation. While we have shown this using the Kodama vector for reference, that is for convenience and is not essential; some other timelike vector field could be used for general spacetime. 

\section{Distinguishing particle creation from other quantum effects}
This section is devoted to studying the discrimination between particle creation and vacuum polarization by choosing suitable coordinates, which cover different parts of the global spacetime. \\

The excellent book by Frolov and Novikov \cite{novikov-book} identifies  (Chapter 10) three relevant quantum effects when discussing black holes: namely (i) particle creation, (ii) vacuum polarization, and (iii) quantum fluctuations of the metric.
As long as we consider a spacetime with curvature smaller than the Planck curvature $\mathcal{R}=\frac{c^3}{\hbar G}\sim 4 * 10^{69} m^{-2}$, we can neglect quantum fluctuations of the metric, so we need only take the other two into account. \\

As is known \cite{novikov-book},  quantum field theory constructed on a curved  space time experiences gravitationally
induced vacuum polarization, which is a deformation of the vacuum fluctuations by the external gravitational field at a given time (similar to the Schwinger deformation by an electric field). Hence, it is
described by 
terms that depend only on the local curvature characterizing the gravitational field at a given location. This effect typically induces a nonzero vacuum expectation value for the stress energy tensor; it is a local effect, in contrast to the non-local nature of particle production. However the vacuum is not a local quantity, it must be determined from global properties; the polarization is local once the vacuum has been chosen. Thus because there are alternative definitions of the vacuum, induced vacuum polarization depends on the vacuum, and there is no unique way to distinguish the particle creation and vacuum polarization contributions to  the total energy-momentum tensor. In this section we discuss how we can sensibly separate quantum particle creation, represented in $\langle T_{\mu \nu} \rangle $, from vacuum polarization, also represented in $\langle T_{\mu \nu} \rangle $ . \\

\textbf{Vacuum polarization and particle creation} At a first glance, it seems that the expectation value for the quantum stress is a local quantity and hence we can straightforwardly describe particle creation by the local quantum stress tensor. 
But finding coordinates that either cover all space time, or that cover only the observer's region, forces us to know 
global properties of the space time (even a particle detector that records the existence of the particle by transition to an excited state has to know about this interaction with the particle along the entire world line, which is non-local \cite{particle detector}). For instance, if one has a coordinate that covers all spacetime in a case when the observer cannot see the black hole interior, the corresponding  vacuum defined in terms of those coordinates will have a mixed state relative to the observer.\\

If observers are sitting outside the black hole, coordinates that only cover the exterior region of the black hole, like tortoise lightcone coordinates $(u,v)$ in the Schwarzschild metric, are suitable coordinate for stress energy  examination. 
Here are some points that help us to specify the vacuum polarization and particle production:
                                         
\begin{itemize}
  \item {If we choose a coordinate (vacuum) that only covers outside the horizon, the quantum expectation value of the stress tensor in this coordinate system only describes vacuum polarization.}
 \item {One test of the vacuum polarization term is that it must be zero in flat regions of space time.}
 \item {Another test for the vacuum polarization term is that it must be zero in local inertial coordinates.}
 \item {For the case of particle creation, there is at least one non-zero component of the quantum stress tensor for all observer in the flat region of the spacetime, provided our vacuum contains the black hole interior region and not only outside the black hole.}
  \item {An important test for having particle creation is that there is a non-zero flux, $\mathcal F$, at the horizon.}
\end{itemize}

\textbf{The two-dimensional general case} We write a 2-dimensional general  metric as 
\be \label{2-dim-metric}
ds^2=-d\xi^+~d\xi^-=-e^{2\rho}dx^+~dx^-
\ee
where $\xi^\pm$ are local null coordinates associated with inertial coordinates, while $x^\pm$ are general null coordinates. Notionally this is a 2-dimensional section of a 4-dimensional spherical spacetime. Here our analysis is for a massless scalar field, which gives the main features of the radiation (in the static case, 81 percent of the radiation is massless \cite{page76}). The general expression for the stress tensor in the 2-dimensional black hole case can be written as (see equation (5.23) in \cite{fabbribook}):
\be \label{stress22}
\langle \psi | T_{\pm\pm}(x) | \psi \rangle =\langle \psi | :T_{\pm\pm}(x): | \psi \rangle +\frac{\hbar}{24\pi}\lbrace \xi^\pm, x^\pm\rbrace
\ee
where the second term is the Schwarzian derivative
\begin{equation}
\lbrace \overline{x}^\pm, x^\pm\rbrace= \frac{d^3\overline{x}^\pm}{d(x^\pm)^3}/ \frac{d\overline{x}^\pm}{dx^\pm} - \frac{3}{2}(\frac{d^2\overline{x}^\pm}{d(x^{\pm})^2}/ \frac{d\overline{x}^\pm}{dx^\pm})^2.
\end{equation}
A direct calculation shows this term 
 is
\be
\label{vacuum-p2}
\frac{\hbar}{24\pi}\lbrace \xi^\pm, x^\pm\rbrace=-\frac{\hbar}{12\pi}(\partial_{\pm}\rho \partial_{\pm}\rho- \partial^2_{\pm}\rho).
\ee
If we go to the case that $x^\pm$ coordinate cover outside the black hole (they are associated with the observer's location) and the $| \psi \rangle $ vacuum covers all space time including the black hole region, the term $\frac{\hbar}{24\pi}\lbrace \xi^\pm, x^\pm\rbrace$ describes vacuum polarization, and the term $\langle \psi | :T_{\pm\pm}(x): | \psi \rangle$ gives the real particle creation part. If $ | \psi \rangle$ is expressed in terms of 
the coordinate $(\bar{x}^+,\bar{x}^-)$, the real particle creation flux relative to a observer with time coordinate $t=\frac{x^- + x^+}{2}$  is equal to
\ba  \label{particle-f2}
\mathcal F &=& \langle \psi | :T_{++}(x): | \psi \rangle - \langle \psi | :T_{--}(x): | \psi \rangle \nonumber\\ &=& -\frac{\hbar}{24 \pi} \left( \frac{d^3\overline{x}^+}{d(x^+)^3}/ \frac{d\overline{x}^+}{dx^+} - \frac{3}{2}(\frac{d^2\overline{x}^+}{d(x^{+})^2}/ \frac{d\overline{x}^+}{dx^+})^2 ~-~ \frac{d^3\overline{x}^-}{d(x^-)^3}/ \frac{d\overline{x}^-}{dx^-} - \frac{3}{2}(\frac{d^2\overline{x}^-}{d(x^{-})^2}/ \frac{d\overline{x}^-}{dx^-})^2 \right)
\ea\\
These equations are general. Their interpretation depends on the vacuum chosen.\\

\textbf{The four dimensional spherical case}
Similarly for a 4-dimensional spherically symmetric black hole we have (see equation (5.157) in \cite{fabbribook} ):
\be \label{stress3}
\langle \psi | T_{\pm\pm}(x) | \psi \rangle =\langle \psi | :T_{\pm\pm}(x): | \psi \rangle -\frac{\hbar}{12\pi}(\partial_{\pm}\rho \partial_{\pm}\rho- \partial^2_{\pm}\rho)+\frac{\hbar}{2\pi} (\partial_{\pm}\rho \partial_{\pm}\phi+ \rho (\partial_{\pm}\phi)^2 )
\ee
where the two-sphere radius is $R =e^{-\phi}$ ($\phi$ is the dilaton field) and
\be \label{metric}
ds^2= -e^{2\rho} dx^+ dx^- + e^{-2\phi}d\Omega^2.
 \ee

Similar to the 2-dimensional case if the coordinates $(x^+,x^-)$ only cover the exterior of the black hole and the $| \psi \rangle $ vacuum covers all space time including the black hole region, the term
 \be \label{vacuum-p4}
 -\frac{\hbar}{12\pi}(\partial_{\pm}\rho \partial_{\pm}\rho- \partial^2_{\pm}\rho)+\frac{\hbar}{2\pi} (\partial_{\pm}\rho \partial_{\pm}\phi+ \rho (\partial_{\pm}\phi)^2 )
  \ee
  is the vacuum polarization and $\langle \psi | :T_{\pm\pm}(x): | \psi \rangle$ is the pure particle creation part.
Therefore the particle flux relative to an observer with time coordinate $t=\frac{x^- + x^+}{2}$  is described by
\be \label{qflux}
\mathcal F=\langle \psi |:T_{++}(x):| \psi \rangle - \langle \psi |:T_{--}(x):| \psi \rangle.
\ee

Covariantly, the created particle energy density and flux \footnote{The  sign convention for the flux was chosen so that the black hole have negative flux.  } are 
\begin{equation}
\rho_p = u^\mu u^\nu \langle \psi|:T_{\mu \nu}:|\psi \rangle, \,\,\,\mathcal F=u^\mu n^\nu \langle \psi|:T_{\mu \nu}:| \psi \rangle
\end{equation} 
respectively where $u^\mu$ is the 4-velocity of the observer and $n^\nu$ is a unit vector normal to $u^\mu$. On choosing the 4-velocity of an observer (detector), we can find the related radiation flux and find the effective temperature for this observer \cite{pady-15}).
\\

\textbf{The Unruh effect}
As regards the Unruh effect, the normal ordered form of the quantum stress tensor, $:T_{uu}:$, for an accelerated observer  in the Minkowski vacuum,  gives the energy that the accelerated detector  absorbs and then makes a transition to the excited state: it is $\langle M|:T_{uu}:|M \rangle= \frac{\hbar a^2}{48 \pi}$. Here the tortoise coordinates $(u,v)$ are the same as the $(-,+)$ coordinates above. This is neither  particle creation nor vacuum polarization, because it is in Minkowski spacetime. It is just an excited state due to acceleration (a "fictitious particle"). \\

\textbf{The Schwarzschild case}
The case of a static black hole is gievn by Schwarzschild space time and its vacuums. There are two well known null coordinates for the Schwarzschild metric. The first is the tortoise lightcone coordinates
\be
u=t-r^*,~~v=t+r^*
\ee
where $r^*=r-2m+2m \ln(\frac{r}{2m}-1)$. These coordinates for $r > 2m$ vary on the range $-\infty < u,v < +\infty$. The second coordinates are  Kruskal-Szekeres null coordinates $(U,V)$
 defined by
\be
U=-4m ~\exp(-\frac{u}{4m}),~~V=4m ~\exp(\frac{v}{4m})
\ee
There are three vacuums for the Schwarzschild metric, which are defined as below \cite{candelas-80}:\\

(i) the \textit{Boulware vacuum} $ |B \rangle$, defined by requiring
normal modes to be positive frequency
with respect to the Killing vector $\frac{\partial}{\partial t}$, with respect
to which the exterior region is static. This vacuum is defined only in the exterior region. The expectation value of stress tensor in this vacuum only describes vacuum polarization.\\

(ii) the\textit{ Unruh vacuum} $ |U \rangle$, defined by taking
modes that are incoming from $\scri^-$ to be positive
frequency with respect to $\frac{\partial}{\partial t}$, while those that
emanate from the past horizon are taken to be
positive frequency with respect to $U$, the canonical
affine parameter on the past horizon. This vacuum is defined on the exterior region together with the interior region associated with particles infalling from that exterior region.   
The expectation value of the stress tensor for this vacuum contains both particle creation and vacuum polarization parts, and has a non-zero flux \cite{visser-97-unruh}.\\

(iii) the \textit{Hartle-Hawking vacuum} $ |H \rangle$, defined
by taking incoming modes to be positive frequency
with respect to $V$, the canonical affine parameter
on the future horizon, and outgoing modes to be
positive frequency with respect to $U$. This covers the entire maximally symmetric vacuum Schwarzschild solution. The expectation value of the stress tensor in this vacuum contains both particle creation and vacuum polarization parts. This has   zero flux (\ref{qflux}), showing this is thermal equilibrium of a black hole with its emitted radiation.\\
\begin{figure}[h]
		\begin{center}
			\includegraphics[scale = 0.5]{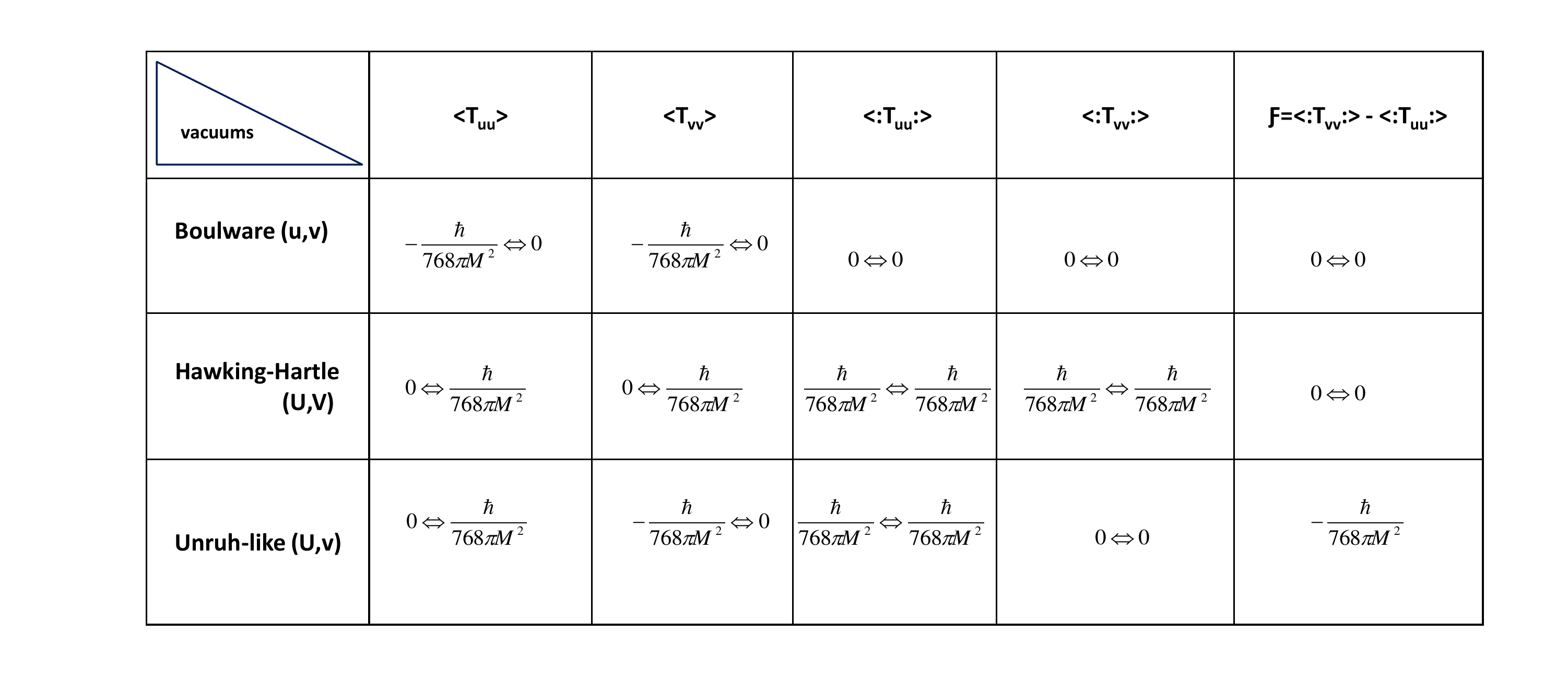}
			\hspace*{10mm} \caption{ \label{sch-table}
				\textit{The expectation value of the energy momentum tensor for different vacuums in 2-dimensional Schwarzschild spacetime. The left number in each column was calculated at the horizon and the right number is at the infinity. The last column represents the radiation flux. By equation (\ref{stress22}), vacuum polarization is given by subtracting the third column from the first, and the second column from the fourth (there is another component not shown here).} }
		\end{center}
	\end{figure}

	A summary of the expectation values of the energy momentum tensor for 2-dimensional Schwarzschild spacetime is shown in the Table \ref{sch-table}.\\
	
 Conceptually, the radiation flux appears for an observer that cannot see parts of the spacetime. Since the Hartle-Hawking vacuum and Unruh vacuum cover inside the horizon, which can not be seen by external observers, the related states for these vacuums are mixed states.

\section{Particle creation in cosmological models}

In this section we investigate particle creation in FLRW metrics using the quantum stress tensor. In particular we discuss de Sitter spacetime in stationary and cosmological coordinates and define the related vacua for these cases. Other aspects of radiation production like the affine parameter approach and the  relation to the inflationary epoch are discussed.\\

Particle creation in cosmological models is substantially different from that in black hole models. The key difference is that in the cosmological case, if event horizons exist for an observer, which will be the case for de Sitter spacetime and for expanding universes with a cosmological constant, every observer can at very late times receive thermal radiation associated with their  cosmological event horizon (note that different observers will have different event horizons). Hence, thermal radiation will in principle appear for observers everywhere. However for the black hole case, there is a unique event horizon for all external observers and only a distant observer outside the black hole will see such radiation at late times. Therefore, we have well-defined observers in the black hole case for whom the thermal temperature and backreaction can be calculated, but in the case of cosmological models, defining the temperature and back reaction of the radiation has ambiguities that must be solved.\\

One should note here that there are various calculations of the vacuum energy of zero point quantum fluctuations, which in principle is one of the main potential candidates for cosmological 
dark energy \cite{weinberg,carroll00}. This zero-point energy of a quantum field diverges, and so is calculated up to a cut-off scale in the case of both flat and curved spacetimes, and turns out to be hugely larger than the observed cosmological constant, which is a major problem for cosmology if this field gravitates (which need not necessarily be the case \cite{Ell13}). This however is an essentially local effect that can occur in cosmology as well as in the black hole case, but is not directly related to either particle creation or vacuum polarization effects, which are confined to curved spacetime backgrounds. We will not consider this 
further here.\\

The standard model in cosmology is based on a homogeneous and isotropic expanding universe. The metric which best describes  the early accelerating expansion (inflation era) and late time acceleration phase (dark energy dominated) is the de Sitter spacetime in non-stationary coordinates, given by the metric (\ref{desitter-nons}).
These coordinates only cover half of the de Sitter manifold $- Z_0^2+Z_1^2+Z_2^2+Z_3^2=\frac{3}{\Lambda},~\Lambda>0$.
However because it is a spacetime of maximal symmetry, one can also use static coordinates with the metric given by (\ref{desitter-sta}). These also only cover part of the de Sitter manifold.
The matter dominated and radiation dominated stages of the universe after inflation are well described by a flat  Friedmann-Lem\'{a}itre-Robertson-Walker (FLRW) model. In contrast to the de Sitter spacetime, this does not have a cosmological event horizon. \\

A full cosmological model will have 
three epochs: an initial inflationary era with metric close to de Sitter, followed by a matter and radiation dominated era, and finally a late time dark energy dominated era that is again like de Sitter. In this section, we discuss particle creation in these various metrics. But one should note the following: the event horizon in cosmology only refers to the far future of the universe. If creation occurs relative to that horizon, it has nothing to do with anything we can observe at the present day.  

\subsection{Particle creation in FLRW spacetime}

The first prominent work to calculate particle creation in a FLRW metric is Parker's work in 1969  \cite{parker-69}, which calculated the probability of finding a particle at the time $t$ according to the Bogolubov coefficients. Being a conformally flat spacetime, one has the ability to specify a well-defined conformal vacuum and determine the mean particle creation rate in conformal time. As discussed above, such non-zero particle creation, which is calculated by comparing the different vacuums according to the Bogolubov coefficients, cannot be considered as pure particle creation: it can be fictitious or quasi-particles. To study the real particle creation rate, one needs to determine that rate 
`from $\langle \psi | T_{\mu \nu} | \psi \rangle $.\\

Using equation (\ref{eq-app.two-stress.1}), it can be shown that the radiation flux (\ref{qflux}) for  the minimally conformally coupled massless scaler field is \emph{zero} for 2-dimensional FLRW models.  For 4-dimensional FRLW metrics, using the equation (\ref{eq-app.four-stress}) to calculate $\langle T_{tx}\rangle$, one can show that \emph{the  radiating particle flux is zero} for conformally coupled massless scalar fields. The non-zero component from (\ref{eq-app.two-stress.1}) and (\ref{eq-app.four-stress}) can be interpreted as vacuum polarization.\\

One can define a cosmological \textit{apparent horizon} as follows. $H$ is a hypersurface in a 4-dimensional spacetime that is foliated
	by 2-surfaces such that $\theta_{(n)}\mid_{H}=0$,
	$\theta_{(\ell)}\mid_{H}\neq0$, and
	$\pounds_{\ell}\theta_{(n)}\mid_{H}\neq0$. 
	This separates regions where either future or past null geodesics going either both inwards or both outwards converge, from where this is not the case. 
	An apparent 
	horizon is
	called \emph{outer} if $\pounds_{\ell}\theta_{(n)}\mid_{H}<0$,
	\emph{inner} if $\pounds_{\ell}\theta_{(n)}\mid_{H}>0$,
 \emph{past} if $\theta_{(\ell)}\mid_{H}<0$, and \emph{future} if
	$\theta_{(\ell)}\mid_{H}>0$. 
	They are trapping horizons, that is, they are related to spacetime regions where particles are causally trapped in the sense that they cannot reach infinity, if they are future directed, but not if they are past directed. Past directed apparent horizons occur in all realistic cosmologies and are related to the existence of the singularity at the start of the universe \cite{ellis-book}.  Future directed apparent trapping horizons occur in cosmology only in universes that recollapse to a second singularity in the future, when they are related to the existence of that singularity. 
	  	The most relevant case in the context of
an ever-accelerating expanding cosmology is the \emph{inner future 
apparent 
horizon} (IFAH) which can occur if the cosmological constant is positive. The physical intuition for this horizon is that it is the boundary of the region so that to the future, 
the cosmological expansion is so strong that even the future directed ingoing null geodesic cannot converge. For de Sitter spacetime this cosmological horizon coincides with the de Sitter event horizon. It is not a trapping horizon because it is future directed; however it can be associated with Hawking radiation.\\

Recently some workers have applied the particle tunneling method to show that there is particle thermal radiation from the  FLRW apparent horizon (cosmological horizon) in the dynamic case \cite{FLRW-radiation}. They have neglected a basic point.  Their calculation is based in the WKB approximation assumed for the light coming from the apparent horizon to the observer. As discussed in \cite{javad-ellis,nielsen-javad}, the WKB approximation cannot hold for light passing the FLRW apparent horizon, except in the case of de Sitter spacetime. As a result, there is no particle creation in the FLRW metric in the matter and radiation dominated eras. 

\subsection{Particle creation in de Sitter spacetime}

Here, we consider particle creation in the stationary and cosmological versions of de Sitter spacetime.

\subsubsection{Particle creation in the stationary de Sitter spacetime}

\begin{figure}[h]
\begin{center}
\includegraphics[scale = 0.5]{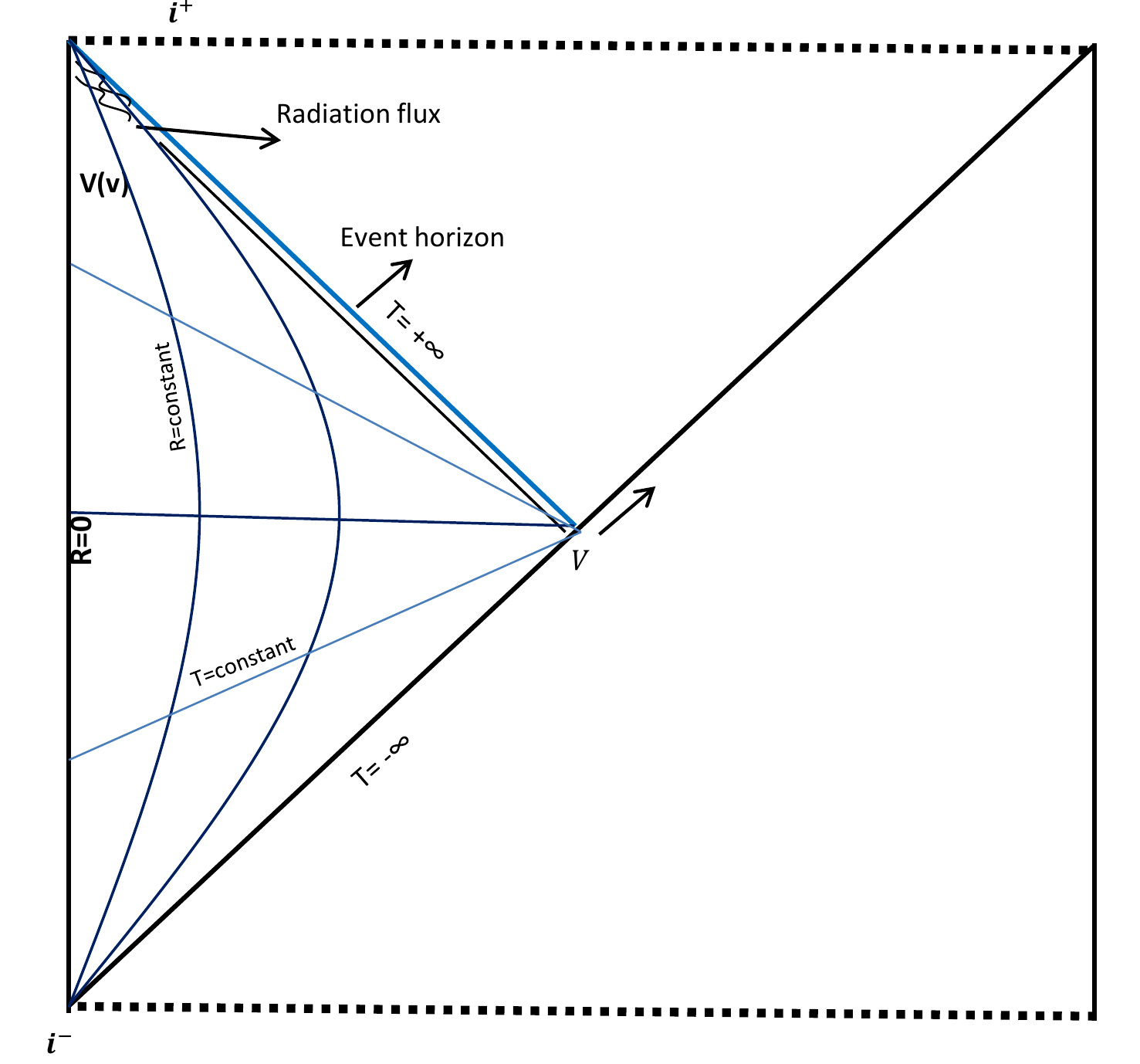}
\hspace*{10mm} \caption{ \label{desitter}
 \textit{Particle creation in the stationary de Sitter spacetime} }
\end{center}
\end{figure}

Historically, by using the path integral method Gibbons and Hawking \cite{Hawking-77} showed that any observer will see thermal radiation in de Sitter spacetime. Let us start our calculation from de Sitter spacetime in stationary coordinates:
\be
\label{desitter-sta}
ds^2 =-(1-\frac{\Lambda}{3}R^2)dT^2+(1-\frac{\Lambda}{3}R^2)^{-1}dR^2 +R^2 d\Omega^2.
\ee
This metric has a time like Killing vector field $\partial/\partial T$, and $R=\sqrt{\frac{3}{\Lambda}}$ is a comoving observer's event horizon with surface gravity $\kappa_c=\sqrt{\frac{\Lambda}{3}}$. These coordinates cover only part of the de Sitter hyperboloid \cite{ellis-book}: they cover the triangle on the left in Fig.(\ref{desitter}) that stretches from $i_-$ to $i_+$.\\

\textbf{The null coordinates}
Defining the new variable $r^*(r)$ by $dr^*=\frac{dR}{(1-\frac{\Lambda}{3}R^2)}$ we get
\be
r^*= \frac{1}{2\sqrt{\frac{\Lambda}{3}}}\ln \left( \frac{1+\sqrt{\frac{\Lambda}{3}}R}{1-\sqrt{\frac{\Lambda}{3}}R}\right).
\ee
Similar to the Schwarzschild coordinate we define tortoise lightcone coordinates $(u,v)$ by 
\be
u=t-r^*,~~v=t+r^*.
\ee
These coordinates are defined for $0 <R< \sqrt{\frac{3}{\Lambda}}$, and vary over the range $-\infty < u,v < +\infty$. The second coordinate definition is Kruskal-Szekeres-like null coordinates $(U,V)$
 given by\footnote{ Note that our definition for $(U,V)$ is the reverse definition of that in the Gibbons paper \cite{Hawking-77}. This choice has been made to be in analogy with the Schwarzschild case where $V=const$ means an ingoing wave and $U=const$ means an outgoing wave.}
 \be
\label{desitter-kruskal}
U=\sqrt{\frac{3}{\Lambda}}\exp(\kappa_c u),~~V=-\sqrt{\frac{3}{\Lambda}}\exp(-\kappa_c v).
\ee
In this coordinate system, the metric becomes
\be
ds^2 =-\frac{4}{(1-\frac{\Lambda}{3}UV)^2} dU~dV +R^2 d\Omega^2.
\ee
As shown in Fig.(\ref{desitter}), light coming from past null infinity, $\scri^-$, and passing near the event horizon is  stretched infinitely. The different with the black hole case is that here we must consider  ingoing null geodesics instead of outgoing ones. The observer coordinates $(u,v)$ cannot see \textit{outside} the event horizon while the $(U,V)$ coordinates cover both outside and inside the horizon. \\

\textbf{The various vacua}
Like the Schwarzschild case, the stationary de Sitter spacetime has three vacua.\\ 

(i) We define the $| B \rangle$ vacuum state (which is like  the Boulware vacuum) for the wave solution in terms of the mode functions $(u,v)$. The corresponding modes are
\be
\frac{1}{\sqrt{4 \pi w}} e^{-iwu},~~~~\frac{1}{\sqrt{4 \pi w}} e^{-iwv}.
\ee
It is not suitable as a vacuum for the stationary 
de Sitter spacetime with particle production because it covers only the inside of the horizon.\\

(ii) 
We define the $| H \rangle$ vacuum state (which is like the Hawking-Hartle vacuum \cite{hawking-hartle}) for the wave solution in terms of the mode functions $(U,V)$. The corresponding modes are
\be
\frac{1}{\sqrt{4 \pi w}} e^{-iwU},~~~~\frac{1}{\sqrt{4 \pi w}} e^{-iwV}.
\ee
It is not suitable as a vacuum for the stationary 
de Sitter spacetime because it covers the inside and outside of the horizon in a time symmetric way. 
It can be shown that 
\begin{equation}
\langle H | :T_{vv}(x): | H \rangle=\langle H | :T_{uu}(x): | H \rangle=\frac{\hbar \Lambda}{144\pi}
\end{equation} at the center, and hence the matter creation flux is zero. It describes a thermal equilibrium universe with temperature $T=\frac{\hbar\kappa_c}{2\pi}$. \\

(iii) We define the $| U \rangle$ vacuum state (which is like  Unruh vacuum ) for the wave solution in terms of the mode functions  $(u,V)$. 
This mode function only covers the upper triangle in the Fig.(\ref{desitter}), 
and it can be a suitable vacuum for particle production in the stationary 
de Sitter spacetime because it covers both inside and outside the horizon in a time asymmetric way. The corresponding modes are
\be
\frac{1}{\sqrt{4 \pi w}} e^{-iwu},~~~~\frac{1}{\sqrt{4 \pi w}} e^{-iwV}.
\ee

As we attribute the $|U \rangle$ vacuum to the $(u,V)$ coordinates, we can calculate the matter creation flux from equation (\ref{particle-f2}): 
\be
\label{flux-desitter}
\langle U | :T_{vv}(x): | U \rangle= \frac{\hbar}{24 \pi} (\frac{\kappa_c^2}{2} )=\frac{\hbar \Lambda}{144\pi}
\ee
where the thermal temperature in this case is 
\begin{equation}\label{eq:temp}
T=\frac{\hbar\kappa_c}{2\pi}.
\end{equation}
 
From the energy conservation point of view (Section II), a Killing horizon exists, the Kodama vector changes sign on the event horizon, and hence energy conservation is valid for the created pair.
\\

\textbf{Affine parameter approach} We can also derive this temperature from the affine null parameter approach \cite{visser-10}. The metric in this case is stationary and the adiabatic condition is satisfied for null geodesics which come from near the event horizon.  Assume that  
we consider the upper triangle in Fig.(\ref{desitter}),  which is covered by the 
		de Sitter space coordinates $(V,u)$. It can be shown that the $v$ parameter is the affine null parameter at future null infinity where the observer receives the radiation, and $V$ coordinate is the past null infinity affine parameter. Therefore, we can write the equation 
\begin{equation}\label{affine_p}
V=p(v)=-\sqrt{\frac{3}{\Lambda}}\exp(-\kappa_c v)
\end{equation}
 for the null geodesic which comes from past null infinity and arrives at the observer's point in future null infinity.  The adiabatic condition will be held for the light passing near the de Sitter space time horizon \cite{nielsen-javad} and it can be shown that the spectrum of the radiation is Planckian \cite{visser-10}. Assuming the adiabatic approximation 
for the null geodesic near the event horizon, we get
\be
T=-\frac{1}{2\pi}\frac{\ddot{p}(v)}{\dot{p}(v)}.
\ee
Using the equation (\ref{desitter-kruskal}) we get the temperature $T=\frac{\kappa_c}{2\pi}$. This temperature is consistent with the temperature (\ref{eq:temp}) from the expectation value of quantum stress tensor (\ref{flux-desitter}). \\

Note however a key feature this derivation makes clear: the observer only experiences particle creation at the end of her history where relation (\ref{affine_p}) diverges, and hence will not experience it at any finite time before then. Hence static de Sitter spacetime regions are not filled with black body radiation everywhere: it comes in to being for an observer only at very late times.\\

\textbf{Vacuum Polarisation} The pure vacuum polarization effect for the 2-dimensional de Sitter spacetime can be calculated in tortoise coordinate $(u,v)$ by the terms 
\be
\langle B |T_{u u}| B \rangle,
\langle B |T_{u v}| B \rangle,
\langle B |T_{v v}| B \rangle,
\ee
where the  $| B \rangle$ state is like the  Boulware state for the tortoise coordinate. The de Sitter metric in the tortoise coordinates is
\be
ds^2=-4\frac{e^{\sqrt{\frac{\Lambda}{3}}(u-v)}}{(1+e^{\sqrt{\frac{\Lambda}{3}}(u-v)})^2}du~dv.
\ee
Here comparing with the equation (\ref{2-dim-metric}) we get $\rho=\ln\left( 2 \frac{e^{\frac{1}{2}\sqrt{\frac{\Lambda}{3}}(u-v)}}{1+e^{\sqrt{\frac{\Lambda}{3}}(u-v)}} \right)$.
According to equation (\ref{vacuum-p2}), the different components of the vacuum polarization are,
\ba
\langle B |T_{u u}| B \rangle &=& -\frac{\hbar}{12\pi}(\partial_{u}\rho \partial_{u}\rho- \partial^2_{u}\rho)\nonumber \\&=&  -\frac{\hbar}{12\pi}\left( \frac{\Lambda}{3}(\frac{1}{2}-\frac{e^{\sqrt{\frac{\Lambda}{3}}(u-v)}}{1+e^{\sqrt{\frac{\Lambda}{3}}(u-v)}})^2 + \frac{\Lambda}{3} (\frac{e^{\frac{1}{2}\sqrt{\frac{\Lambda}{3}}(u-v)}}{1+e^{\sqrt{\frac{\Lambda}{3}}(u-v)}})^2 \right)= -\frac{\hbar \Lambda}{144\pi}
\ea
\ba
\langle B |T_{v v}| B \rangle &=& -\frac{\hbar}{12\pi}(\partial_{v}\rho \partial_{v}\rho- \partial^2_{v}\rho)\nonumber \\ &=& -\frac{\hbar}{12\pi}\left( \frac{\Lambda}{3}(\frac{1}{2}-\frac{e^{\sqrt{\frac{\Lambda}{3}}(u-v)}}{1+e^{\sqrt{\frac{\Lambda}{3}}(u-v)}})^2 + \frac{\Lambda}{3} (\frac{e^{\frac{1}{2}\sqrt{\frac{\Lambda}{3}}(u-v)}}{1+e^{\sqrt{\frac{\Lambda}{3}}(u-v)}})^2 \right) = -\frac{\hbar \Lambda}{144\pi}
\ea
\ba
\langle B |T_{u v}| B \rangle &=& -\frac{\hbar}{12\pi}(\partial_{u}\rho \partial_{v}\rho- \partial^2_{uv}\rho) \nonumber \\ &=& -\frac{\hbar}{12\pi}\left( \frac{-\Lambda}{3}(\frac{1}{2}-\frac{e^{\sqrt{\frac{\Lambda}{3}}(u-v)}}{1+e^{\sqrt{\frac{\Lambda}{3}}(u-v)}})^2 - \frac{\Lambda}{3} (\frac{e^{\frac{1}{2}\sqrt{\frac{\Lambda}{3}}(u-v)}}{1+e^{\sqrt{\frac{\Lambda}{3}}(u-v)}})^2 \right)= +\frac{\hbar \Lambda}{144\pi}
\ea
These equations show that the vacuum polarization flux relative to a observer with time coordinate $t=\frac{u + v}{2}$  is zero. The vacuum polarization is homogeneous for the de Sitter spacetime, and it reduces to zero in the Minkowski limit $\Lambda\rightarrow 0$, which  show these results are due to gravitational effects in quantum field theory. The energy density of the vacuum polarization for the observer who is sitting in the center $r=0 \Leftrightarrow u-v=0$ is zero: 
\begin{equation}
\rho_{vp}=\langle B |T_{u u}| B \rangle+\langle B |T_{v v}| B \rangle+2\langle B |T_{u v}| B \rangle = 0.
\end{equation}\\

\begin{figure}[h]
\begin{center}
\includegraphics[scale = 0.5]{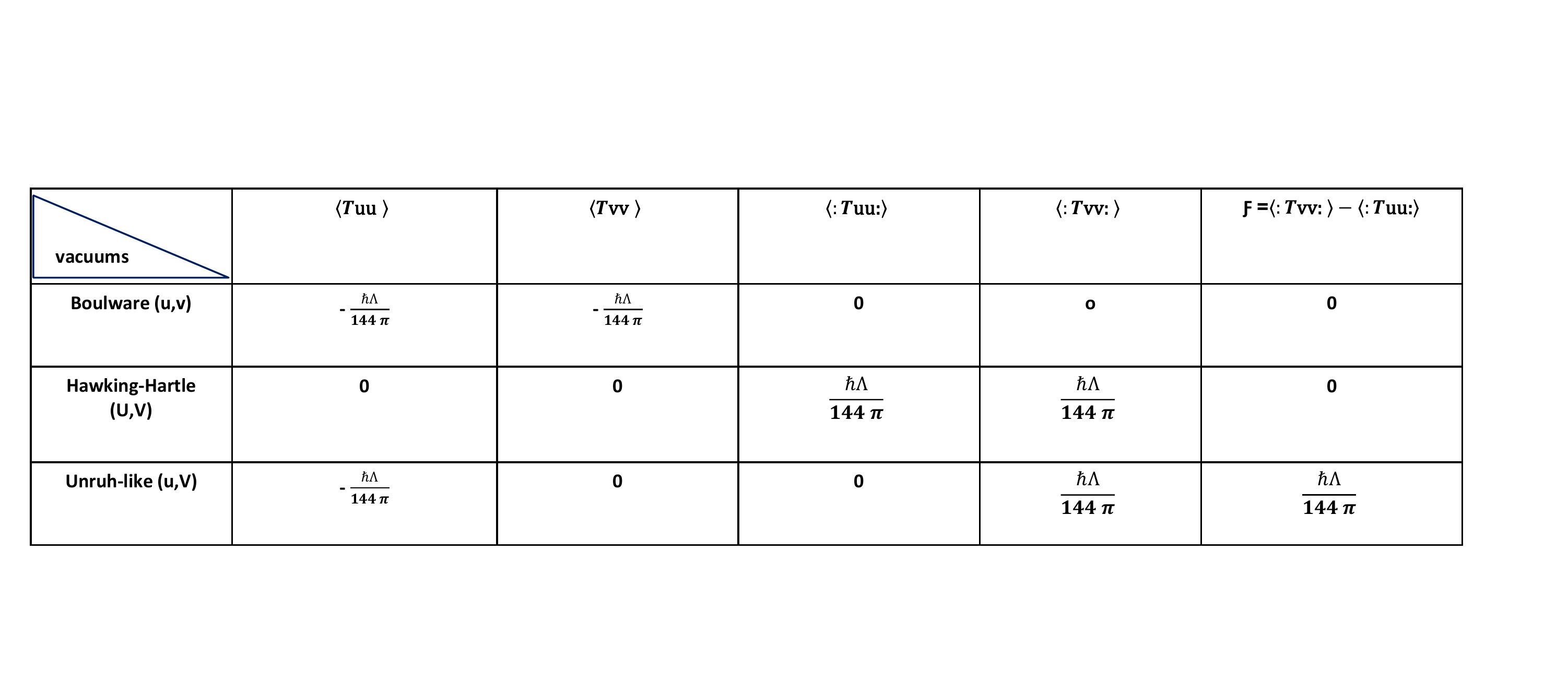}
\hspace*{10mm} \caption{ \label{desitter-table}
 \textit{The expectation value of the energy momentum tensor in different vacuum for 2-dimensional de Sitter spacetime observed at the center. The meaning is as in the case of Fig.(\ref{sch-table}).} }
\end{center}
\end{figure}

\textbf{The total effects}
A summary of the expectation values of the energy momentum tensor for 2-dimensional static de Sitter spacetime is given in  Fig.(\ref{desitter-table}).  Here we see that particle creation flux occurs only for the Unruh-like vacuum and the expectation value for the Boulware vacuum state represents pure vacuum polarization.

\subsubsection{Particle creation in the cosmological de Sitter spacetime}

The cosmological de Sitter spacetime is described by the metric 
\be
\label{desitter-nons}
ds^2=-dt^2+e^{2\sqrt{\frac{\Lambda}{3}}t}(dr^2+r^2 d\Omega^2).
\ee
These coordinates only cover half the de Sitter manifold $- Z_0^2+Z_1^2+Z_2^2+Z_3^2=\frac{3}{\Lambda},~\Lambda>0$; consequently 
as shown in Fig.(\ref{desitter-cos}), this spacetime covers the upper half of the total de Sitter manifold. Like the case of stationary coordinates, this spacetime has an event horizon at $r=\sqrt{\frac{3}{\Lambda}}e^{-\sqrt{\frac{\Lambda}{3}} t}$ and the $(t,r)$ coordinates cover both inside and outside the horizon. Using the coordinate transformation,
\be
R= e^{\sqrt{\frac{\Lambda}{3}}t} r,~~T=t+\int \frac{\sqrt{\frac{\Lambda}{3}} R}{1-\frac{\Lambda}{3}R^2} dR=t-\frac{3}{2\Lambda}\ln{(1-\frac{\Lambda}{3}R^2)}
\ee
we can relate stationary de Sitter coordinates $(T,R)$ with metric (\ref{desitter-sta}) to the cosmological coordinates $(t,r)$ with metric  (\ref{desitter-nons}). 

This coordinate transformation  changes the particle creation effects discussed above in the static case for an observer who is located at $r=0$ in the expanding case. Particle creation for the cosmological de Sitter spacetime can be seen by other methods as well \cite{pady-three stage, vanzo-09}. Again the effect only occurs at late times when the affine parameter relation diverges. It does not manifest for an observer at a finite time in her history (because no matter what that finite time is, there is still an infinite time to run before the event horizon is reached).\\

\textbf{Vacuum Polarization} Using equation (\ref{eq-app.two-stress.1}), the expectation value of the quantum stress tensor for a general FLRW metric is 
\ba
\langle T_{uu} \rangle = \langle T_{vv} \rangle= \frac{\hbar (a \ddot{a}-\dot{a}^2)}{48 \pi}, \nonumber \\
\langle T_{uv} \rangle= \frac{-\hbar (a \ddot{a})}{48 \pi}.
\ea
This shows that the FLRW model has no flux of  Hawking radiation and no vacuum polarization. Thus in the case of de Sitter spacetime we get
\be
\langle T_{uu} \rangle = \langle T_{vv} \rangle= 0,~~\langle T_{uv} \rangle= \frac{-\hbar \Lambda e^{2\sqrt{\frac{\Lambda}{3}}t}}{144 \pi}
\ee
We can calculate the energy density of vacuum polarization(VP) for the comoving observer $u^\mu=(1,0)$ inside the horizon 
as:
\be
\rho_{vp}=\frac{1}{a^2}(\langle \overline{x}^\pm| T_{uu} |\overline{x}^\pm \rangle+\langle \overline{x}^\pm| T_{vv} |\overline{x}^\pm \rangle+2\langle \overline{x}^\pm| T_{uv} |\overline{x}^\pm \rangle)=\frac{-\hbar \Lambda}{72\pi}.
\ee
Like in the case of stationary coordinates, the vacuum polarization in a cosmological de Sitter spacetime is homogenous. The comoving cosmological observer will see a constant vacuum polarization energy density which is smaller in magnitude than the cosmic fluid by order $\hbar$ .\\

The Kodama vector (\ref{kodama}) 
changes sign on the 
event horizon, which is a null surface, so particles and anti-particles have different signs for their energy and hence energy conservation is obeyed for any particles created there; hence particle creation is possible. One might think that such particles created at the event horizon are diluted away by cosmological infinite redshift, $1+Z_c=\frac{e^{\sqrt{\frac{\Lambda}{3}}t_0}}{e^{\sqrt{\frac{\Lambda}{3}}t_{\infty}}}$, but this argument is not true. This is like the black hole case where the created particles near the horizon get infinitely redshifted when it reaches future infinity. 

\subsubsection{The expanding de Sitter era in standard cosmology}

\begin{figure}[h]
	\begin{center}
		\includegraphics[scale = 0.5]{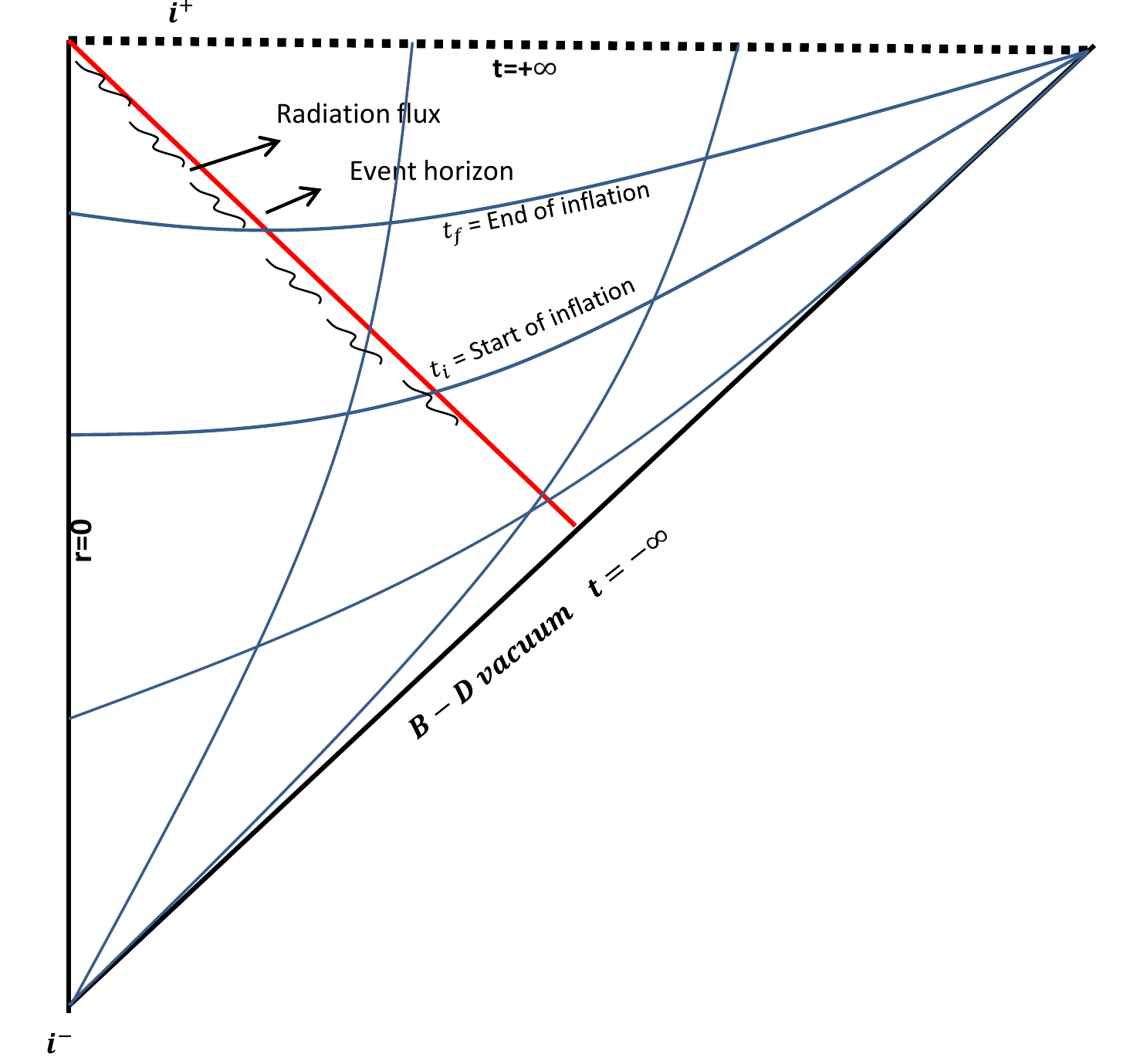}
		\hspace*{10mm} \caption{ \label{desitter-cos}
			\textit{Particle creation in the cosmological de Sitter spacetime} }
	\end{center}
\end{figure}

The standard model of cosmology uses the expanding de Sitter  metric (\ref{desitter-nons}) for describing both the early inflationary era and the late dark energy dominated era, but with different values of the constant $\Lambda$ in these two phases, which are separated by a matter and radiation dominated era. Since there is both a start time and an end time for the inflationary era, this phase of the universe's history  is only covered by the small part of the de Sitter spacetime which is located between times $t_i$ and $t_f$ in  Fig.(\ref{desitter-cos}). The late dark energy dominated phase starts at a time $t_{DE}$ and (assuming it is a cosmological constant) lasts forever, and so has an associated event horizon as $t \rightarrow \infty$. \\
  
\textbf{Late time particle creation} Observers have an event horizon when vacuum energy dominates at late times (so future infinity is spacelike) so one might expect Hawking radiation to be associated with this horizon at very late times.  Note that the  event horizon for late time accelerating phase is the same as the cosmological apparent horizon. Overall, the initial state (the vacuum choice at the start) and state at the event horizon of the universe need knowledge of the pre-inflation era and final fate of our universe respectively. If we have a finite time for an exponentially accelerating era, it is hard finding the particle creation term from the quantum stress tensor when we do not know the initial state. We can however use the tunneling method \cite{vanzo-09} to calculate the flux of radiation at very late times. However since any particles created near the cosmological event horizon would arrive at the center at very late times, in practice a comoving cosmological observer cannot see any associated created particles at the present time in an expanding universe. They will arrive at her world line only in the very far future. 
Thus particle creation in the late dark energy dominated phase will have no present day observational effects in cosmology, because the event horizon for any observer only exists in the far future of the universe. 
\\

\textbf{Inflationary particle creation}  Note that the de Sitter horizon for the inflationary era is not a cosmological event horizon, because the inflationary era comes to an end and is replaced by radiation dominated era. There  is however a cosmological apparent horizon (an anti trapping surface), and this can lead to particle creation (because Hawking radiation can be associated with apparent horizons rather than event horizons \cite{visser-03}).\\

To study particle creation in the inflationary era, we have to notice these points: 
\begin{itemize}
 \item The standard choice of the vacuum state is the Minkowski vacuum of a comoving observer in the far past  $t=-\infty$ (when all comoving scales were far inside the Hubble horizon), which is called the \textit{Bunch-Davies} (B-D vacuum), and is shown in  Fig.(\ref{desitter-cos}). There are some discussions about suitability of choosing this vacuum \cite{non-bunch-davies} because,  
 it is defined at past time infinity, which is not applicable to standard big-bang cosmology where inflation started a finite time ago. Changing the vacuum will result in a different power spectrum of the quantum fluctuations and expectation value of the stress tensor. For example as we saw in the last subsection, if we choose another vacuum like the vacuum related to the maximally extended de Sitter spacetime, there is no matter creation flux. Even observational parameters like the primordial power spectrum and bispectrum in the CMB change on choosing another vacuum \cite{non-bd-cmb}. To know the correct vacuum, we have to know the pre-inflationary models to calculate the initial state. Since the initial state of the big bang cosmology is directly connected to quantum gravity problems, we cannot know the initial state at present. It seems that effective field theory can be a tool to estimate the effective initial state for pre-inflationary models. \\
\item The un-renormalized expectation value for the stress tensor has ultraviolet (UV) divergences in the B-D vacuum which cannot be renormalized by the usual counterterm. Recently, people have applied different renormalization methods for UV divergences in cosmological models \cite{uv-renormalization}. This UV divergency comes from choosing the B-D vacuum, or is generated by the unphysical procedure of sudden matching between different cosmological eras \cite{stress-bunch-davies}.

\item The expectation value of the quantum stress tensor in the B-D vacuum according to the observable coordinates describes vacuum polarization, particle creation, and vacuum energy effects (in the case that we consider background wave number renormalized by the cut off), see \cite{stress-bunch-davies}. As shown in Fig.(\ref{desitter-cos}), choosing a vacuum like the B-D vacuum, which represents the initial state of the field at far past infinity (deep inside horizon), generates  a particle creation term in the expectation value of the field in inflationary models. However inflation starts from an initial time, not from infinity.\\
\item When we change coordinates from the cosmological frame to the stationary frame, the cosmological redshift changes to a gravitational redshift because of that transformation. The point is that the attributed temperature (or wave frequency) is calculated by comparing the vacuum at future and past infinities, giving the temperature at the observer's point in the future. To see this, consider the Kodama vector given by equation (\ref{kodama}). For the stationary de Sitter spacetime this is $ (-\frac{2}{3}\Lambda R,0,0,0)$ and for the cosmological de Sitter spacetime it is $(1,-\sqrt{\frac{\Lambda}{3}}r,0,0)$). One can argue that Kodama observers who lie inside the horizon and have 4-velocity vector $\hat{K}=\frac{K}{|K|}$ see the divergence in frequency measured by such observers at the horizon $\hat{\nu}=\frac{\nu}{|K|}$ . Therefore, such observers measure a thermal spectrum with temperature $\hat{T}=\frac{T}{|K|}$ (see \cite{javad-radiation} for more discussion) which has a divergence at the cosmological apparent horizon.

\item An applicable method for this era is the tunnelling method. The advantage of this method is that we do not need to know the initial state of the pre-inflation era or the global event horizon. We just need the WKB approximation \cite{nielsen-javad} for ingoing waves near the de Sitter  horizon (the cosmological apparent horizon). In contrast to the black hole case where we have a apparent horizon for outgoing null geodesics, which is a trapping horizon, here we have a cosmological apparent horizon for ingoing null geodesics, which is an anti-trapping horizon (it is a time reversed trapping horizon). In black hole cases, outgoing null geodesics are received by the observer; here ingoing null geodesics are received by the observer. For the case of a de Sitter era that comes to end we have a finite burst of  Hawking radiation flux  associated with the apparent horizon which reaches the observer at very late times and is diluted by cosmological expansion by a factor $(\frac{1}{1+z})^2$ (the reason for power 2 is that  one factor $\frac{1}{1+z}$ is due to the cosmological Doppler redshift of the photon and second is because of the arrival rate of the photons is also reduced by the same factor) where $z$ is cosmological redshift. The resulting Planckian spectral flux is
\be
\rho(w)=\frac{dw}{2 \pi} \frac{\Gamma(w)}{e^{\frac{2\pi w}{\kappa_c}}-1}
\ee 
where $\Gamma(w)$ is the frequency-dependent (graybody)
transmission coefficient for the outgoing particle to
reach future infinity without backscattering. The total luminosity for this radiation is $L=\int_{0}^{+\infty} \rho(w)$. As a result of the redshift factors, this  particle creation term does not give a significant contribution to the total matter in the universe or to its expansion rate. Nor does it affect any observations we might make at the present time.\\

\end{itemize}

But in any case, one does not need any concept of horizons or particle pair creation in order to derive the quantum fluctuations that lead to structure formation in an inflationary universe: see e.g. ~\cite{Muk05}. The basis of inflationary model perturbations is a mechanism that generates quantum fluctuation of the matter field because of the non-commutativity of the field state and momentum; these become classical after horizon exit and then  become seeds for cosmological structure formation \cite{Muk05}.

\section{Different aspects of semi-classical quantum effect for compact star and black hole}
In this section, we discuss different aspects of quantum vacuum, like vacuum polarization, for   compact stars and asymptotically flat spacetimes. For the black hole case we consider the effect of applying the adiabatic condition on the quantum stress tensor, looking at the spectrum of negative energy particles  which fall into the black hole and considering the back-scattering of radiation.

\subsection{Quantum stress tensor for compact stars}

If the spacetime does not possess an event or apparent horizon (the case of stars and planets) then
the above complications go away and we only have one vacuum state to deal 
with; the Boulware vacuum, which encompasses 
the entire 
space. There is no event horizon, and so there are no processes associated with an event horizon.\\

The vacuum solution around the compact star can be described by a Schwarzschild exterior solution in tortoise  coordinates:
\be
ds^2=-(1-\frac{2m}{r})du~dv+r^2 d\Omega^2
\ee
where $r>2m$. In these coordinates, the vacuum polarization for the compact star is the same term $\langle B| T_{\mu \nu}|B \rangle$ as discussed before \cite{visser-97-boulware}.
One can take the vacuum solution before the star collapses, $|in\rangle$, and calculate the quantum stress tensor after the star collapses according to this prescription. On the other hand we can read the Bogoliubov coefficients and find the particle number operator, $\langle in|N_i|in\rangle=\langle in|a_i^{out~\dag}a_i^{out}|in\rangle=\sum_j |B_{ij}|^2$. Both of these methods calculate the same vacuum polarization effect (not particle creation). Note that Fabbri \emph{et al.} \cite{Fabbri:2004} have shown that the particle number operator is proportional to total quantum energy in 2-dimensional space time.\\

As an example, consider a static star with spherically symmetric distribution of perfect fluid. The line-element can be written,
\be
ds^2= -e^{2\alpha(r)} dt^2 +e^{2\beta(r)}dr^2+ r^2 d\Omega^2.
\ee
One can define double null coordinates $(u,v)$ for this metric 
by $du=dt-dr^*$ and $dv=dt+dr^*$ where $dr^*=e^{\beta(r)-\alpha(r)}dr$. For these coordinates,  or any other null coordinate $(U(u),V(v))$,  we have $\frac{\partial u}{\partial r}=-\frac{\partial v}{\partial r}$. Therefore from the equations (\ref{vacuum-p2}) and  (\ref{vacuum-p4}) the vacuum polarization components give 
the equation $\langle T_{uu}(x)\rangle= \langle T_{vv}(x)\rangle$. Therefore, like in the Schwarzschild case, the flux of vacuum polarization for a comoving observer is zero. As a result, in the static spherically symmetric metric the general matter flux relative to the comoving observer, given by $\mathcal F=\mathcal F_{VP} + \mathcal F_{PC}$, is equal to only the particle creation flux, i.e. $\mathcal F_{VP}=0$.  But there is no particle creation flux as there is no event horizon for compact stars so $\mathcal F_{PC}=0 \Rightarrow \mathcal F=0$.\\

If a central object in asymptotically flat spacetime is sufficiently non-compact, $2m \ll r$, then the spacetime is closely Minkowskian and there are no horizons.In this case the quantum vacuum polarization effect is proportional to the two factors $\hbar$ and $\frac{2m}{r}$ \cite{Fabbri:2004, visser-97-boulware}. The first factor shows that this effect is a small quantum correction. The second factor shows that such a vacuum state will be virtually indistinguishable
from the Minkowski vacuum state. Hence, the expectation value of the (renormalized) quantum stress tensor will be negligible throughout the entire
spacetime. In other words, this case is the weak field limit, and since the linear relation between two coordinates $\xi^\pm$ and $x^\pm$ is dominant in the equations (\ref{vacuum-p2}) and (\ref{vacuum-p4}), 
the  vacuum polarization term is zero.

\subsection{Black Hole Spacetimes}
Here we want to find the trace of the adiabatic condition \cite{visser-10} in the quantum stress tensor when black holes occur. Let the affine parameters on the null generators of
past and future null infinity, $\scri^-$  and $\scri^+$, be $U$ and $u$. Null geodesics 
from past null infinity pass through the center and reflect off the center at $r = 0$. We express the relation between these two affine parameters provided by 
the null curves as $U=p(u)$ and define the function
\be
\kappa=-\frac{\ddot{p}(u)}{\dot{p}(u)}.
\ee
where $\dot{}=\frac{d}{du}$. With this definition we can show that $\frac{\dddot{p}(u)}{\dot{p}(u)}=-\dot{\kappa}+\kappa^2$. The non-zero term in the particle flux equation (\ref{particle-f2}) in the two dimensional case becomes
\be
\langle \psi | :T_{--}(x): | \psi \rangle= \frac{\hbar}{24 \pi} (\dot{\kappa}+\frac{\kappa^2}{2} )
\ee
where $U=\overline{x}^-$ and $u=x^-$. Now assuming the adiabatic condition $|\dot{\kappa}(u_*)|\ll \kappa(u_*)^2$ along the $u_*$ reference null curve completely traversing the body, we get a non-zero term for particle flux.
 For the stationary case $\kappa=\frac{1}{4M}$, so we get the standard thermal flux 
 \begin{equation}
 \langle\psi | :T_{--}(x): | \psi \rangle= \frac{\hbar}{768 \pi M^2}.
 \end{equation}

 This calculation can be extended to 4-dimensional spacetime using the equation (\ref{qflux}) (equation (5.155) in \cite{fabbribook}). 
As discussed in \cite{visser-10}, the adiabatic condition (eikonal approximation) is equivalent to the statement that a photon emitted near the peak of the Planckian spectrum should not see a large fractional change in the peak energy of the spectrum over one oscillation of the electromagnetic field (that is, the change in spacetime geometry is adiabatic as seen by a photon near the peak of the Hawking spectrum). This condition on the tunneling method for black holes radiation gives a constraint on black holes that can emit radiation \cite{javad-ellis}.

\subsection{Black hole radiation from quantum stress tensor}
Let us discuss about the following important question:

\begin{quote}
\emph{Can we recognize particle creation from the local character of the quantum stress tensor, or do we need some global information about the spacetime?}
\end{quote}

Generally the final fate of gravitational collapse is a black hole or compact star. For compact stars,  only the vacuum polarization contributes to the quantum stress. For asymptotically flat or conformally flat space time like {FLRW}, since the affine parameters $U_{in}$ (we drop the in index in the formulas) and $u$ on the null generators of past and future null infinity are linearly related, the particle creation flux part is zero from the equation (\ref{particle-f2}). But gravitational collapse that has its final fate as a black hole has two affine parameters exponentially related near the horizon. This exponential relation gives a non-zero term for the created matter flux.\\

\begin{figure}[h]
\begin{center}
\includegraphics[scale = 0.5]{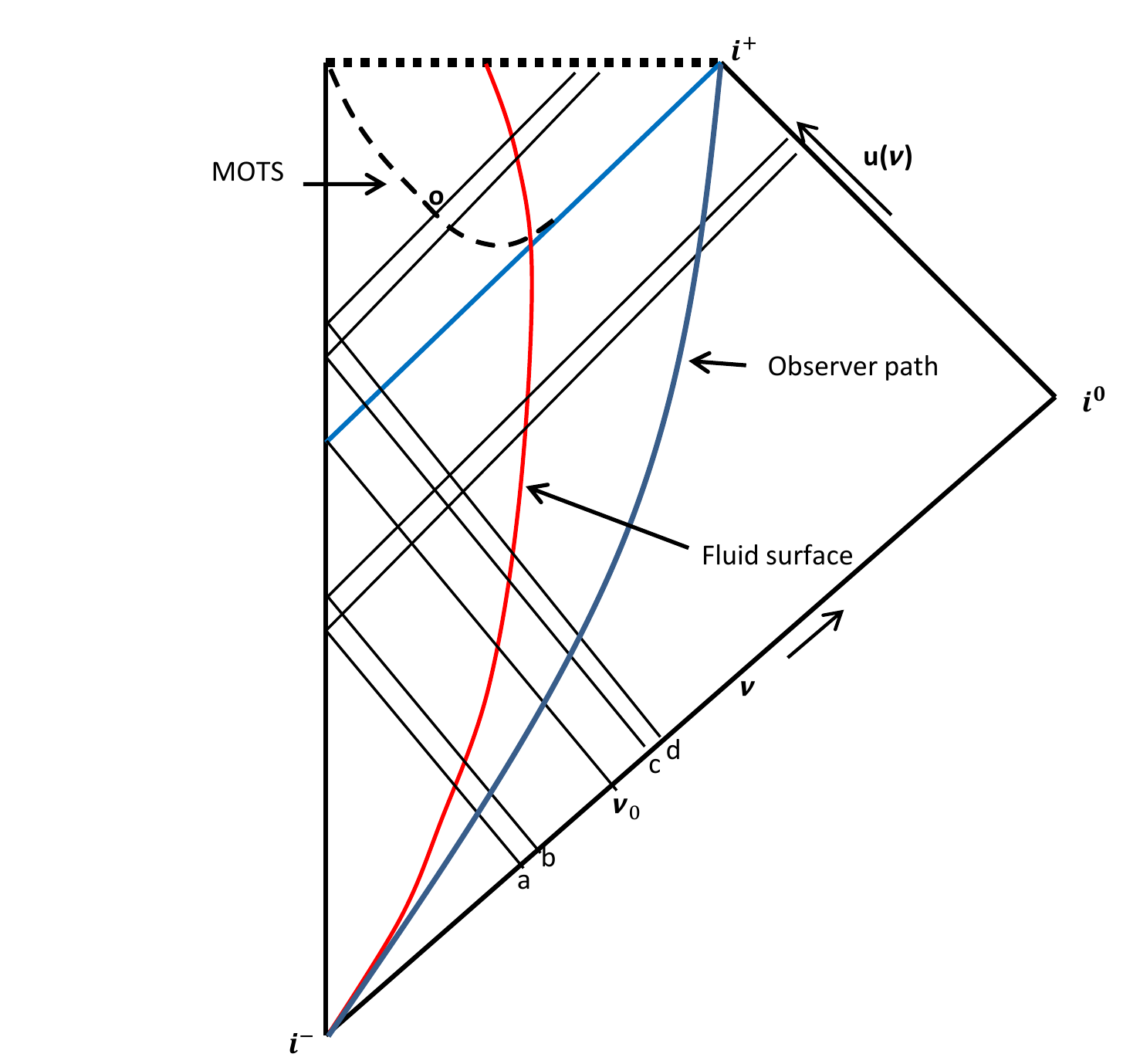}
\hspace*{10mm} \caption{ \label{penrose-BH}
 \emph{Null trajectory for the spherical black hole collapse} }
\end{center}
\end{figure}

\textbf{Collapse to a Schwarzschild black hole: Event Horizons}  For more clarification, consider collapse to a  black hole so that a Schwarzschild black hole forms finally. Following the calculation in \cite{parker-book} (equation (4.24)), if $\lambda$ is an affine parameter which separates outgoing null geodesic at future null infinity and ingoing null geodesics at past null infinity, and $u=t-r^*$,  we get
\be
u=2E\lambda-4m \ln(\frac{\lambda}{k_1})
\ee
at future infinity and
\be
U - U_0 = -k_2 \lambda
\ee
at past infinity ($k_1$ and $k_2$ are constants).  Therefore, we can write
\be \label{affine-sch}
u=-\frac{2E}{k_2}(U - U_0)-4m \ln(-\frac{U - U_0}{k_1 k_2})
\ee
The quantity $U_0$ determines the last ingoing null ray that can come out of the black hole, as shown in Fig.(\ref{penrose-BH}). Far from the horizon, the relation between two affine parameter is linear, and according to relation (\ref{particle}) there is no particle creation. In addition, in the zero-limit for the black hole mass we have zero particle creation. Only near the black hole horizon do we get the exponential relation between parameters and on using the equation (\ref{particle}) we get a non-zero particle flux. Using the affine relation (\ref{affine-sch}), we get the Bogolubov coefficients :
\be
\alpha_{ww'}=-C \int_{-\infty}^0 ds (\frac{w'}{w})^{\frac{1}{2}} e^{i(w'-w\frac{2E}{k_2})s} e^{iw'U_0} \exp[iw4m \ln(s/k_1k_2)]
\ee
and
\be
\beta_{ww'}=-C \int_{-\infty}^0 ds (\frac{w'}{w})^{\frac{1}{2}} e^{i(w'+w\frac{2E}{k_2})s} e^{iw'U_0} \exp[iw4m \ln(-s/k_1k_2)]
\ee
(this is an extension of equations (4.75) and (4.76) in \cite{parker-book}).\\

In the case $|\alpha_{ww'}|^2 \neq\exp(8\pi m w) |\beta_{ww'}|^2$ we have deviation from thermal radiation at a distance from the horizon.
All observer paths will end at $i^+$ as shown in Fig.(\ref{penrose-BH}). Hence, all observers will see thermal radiation coming from near the horizon at late times. Note that all the above calculation are related to the case when we know the final (global) scenario to be collapse to a black hole  and can use  equation (\ref{affine-sch}) for the Schwarzschild geometry. If the final fate of the collapse is a star, the linear relation between affine parameters  gives no particle creation. We are also here assuming no infalling matter or radiation. Note that one can calculate the expectation value of the stress tensor in terms of the past null infinity affine parameters $(U_{in},V_{in})$, which are related to the future affine parameter by equation (\ref{affine-sch}). This vacuum state is called the "in" vacuum state,  and the resulting expectation values of stress tensor are calculated for the Schwarzschild metric in the chapter 5 of the book \cite{fabbribook}.\\

\textbf{Collapse to a Schwarzschild black hole: Apparent Horizons} For calculating particle creation at the point $O$ on the apparent horizon (MOTS), we need to consider two ingoing null rays $c$ and $d$ and their  affine distance at past null infinity and at the future singularity (since these two null rays end at the singularity). We compare the affine distance of these two null rays before falling into the singularity with their affine distance at past null infinity. Since we are inside the event horizon the $t$ and $r$ coordinate change their character. We define new coordinates $\frac{dr^*}{dr}=-\left( \frac{2m}{r}-1\right)^{-1}$ and $u'=r^*-t$. Then the affine distance relation gives
\be
u'=2E\lambda+4m \ln(\frac{\lambda}{k_1})
\ee
at future infinity where $E<0$ and $k_1,\lambda >0$ and
\be
U - U_0 = k_2 \lambda
\ee
at past infinity where $k_2 >0$.  Therefore, we can write
\be
u=\frac{2E}{k_2}(U - U_0)+4m \ln(-\frac{U - U_0}{k_1 k_2}).
\ee
We see that similar relations will hold between the affine parameters. With the same reasoning as in the case of outside the horizon, the second exponential part gives a non zero term for particle creation in equation (\ref{particle}), which is significant for outgoing null rays which pass near the event horizon. The important point is that the $-$ sign behind the second term in the above equation causes that the thermal radiation near the horizon but inside the horizon has negative energy.
 Calculation of the Bogoliubov coefficient near the horizon and inside the black hole gives  
\be
|\alpha_{ww'}|^2 = e^{-8\pi m w} |\beta_{ww'}|^2
\ee	
Hence the number of created particle at late time near and inside the horizon is
\be
\langle N \rangle = \frac{\Gamma(w)}{ e^{-8\pi m w}-1}
\ee	where $\Gamma(w)$ is backscattering factor for wave packet. \\

Since the expectation value of the quantum stress tensor has an internal integration over the wave frequency, we cannot read the spectrum of the radiation from it. As is discussed in \cite{visser-10}, the adiabatic condition is the necessary physical condition for having a Planckian spectrum. For a collapsing model that terminates in a Schwarzschild black hole, the Planckian spectrum comes from the logarithmic term (\ref{affine-sch}) which means that only waves that pass near the event horizon have a Planckian spectrum. 

\subsection{An example: two dimensional Schwarzschild black holes}
As is  known,  the tensor $\tmn{\mu \nu}$ has two advantages in determining particle creation effects: the first is that it gives a covariant definition of real particle creation , and the second is that if it becomes zero in one coordinate system, it remains  zero in all other coordinates.  To calculate the quantum stress tensor in the Openheimer-Snyder collapse case, consider the  interior FLRW metric as
\ba
 ds_{\rm F}^2 = - dt^2 + a(t)^2 \left(\, \frac{dr^2}{1 - k r^2} + r^2\, d\Omega^2 \,\right) \, ,
\label{eq-ES.bg-frw}
\ea
where $a(t)$ is the scale factor, $k$ is the spatial curvature, $r$ is the comoving radius, and $t$ is the proper time of the comoving observer, that is, the cosmological time. The double null form of the metric is given by
\be
   ds_{\rm F}^2 = - a^2 du\, dv + R^2\, d\Omega^2 \, ,
\label{eq-ES.bg-frw.null}
\ee
where
\begin{subequations}
\ba
 d\eta &\defeq& \frac{dt}{a} \, , \quad d\chi \defeq \frac{dr}{\sqrt{1 - k r^2}} \, ,
\label{eq-ES.bg-frw.eta.chi} \\
 u &\defeq& \eta - \chi \, , \quad   v \defeq \eta + \chi \, .
\label{eq-ES.bg-frw.u.v}
\ea
\end{subequations}
Assume the light which comes from past null infinity where the metric is Minkowski space time passes  through the center and reaches a point inside the FLRW space time.
Let $u=x^-$ and $v=x^+$ be a null coordinate inside the FLRW part and  $\bu=\overline{x}^-$ and $\bv=\overline{x}^+$ be initial null coordinates at past null infinity which we denote by$|x^\pm \rangle$ and $|\overline{x}^\pm \rangle $ . As shown in \cite{fabbribook}
\be \label{stress2}
\langle \overline{x}^\pm| T_{\pm\pm} |\overline{x}^\pm \rangle =\langle x^\pm| T_{\pm\pm}|x^\pm \rangle-\frac{\hbar}{24\pi}\lbrace \overline{x}^\pm, x^\pm\rbrace
\ee

The observer's radiation flux at the observation moment $\eta_0$ is given by
\be
\langle \overline{x}^\pm| T_\eta^\chi  |\overline{x}^\pm \rangle = \frac{1}{a_0^2}\langle \overline{x}^\pm|T_{\eta\chi} |\overline{x}^\pm \rangle  = \frac{1}{a_0^2}\langle \overline{x}^\pm|( T_{uu}-T_{vv}) |\overline{x}^\pm \rangle
\ee
Substituting equation (\ref{stress2}) into the above equation one gets
\be
\langle \overline{x}^\pm| T_\eta^\chi  |\overline{x}^\pm \rangle =  \frac{\hbar}{24 \pi a_0^2} \left(  \frac{3}{2}(\frac{d^2\overline{u}}{du^2}/ \frac{d\overline{u}}{du})^2 -\frac{d^3\overline{u}}{du^3}/ \frac{d\overline{u}}{du}\right)
\ee
Without giving the details of the calculation, the $\overline{u}$ which comes from the Minkowski part at past null infinity and passes through the center to arrive at an IMOTS (the Inner Marginally Trapped 3-Surface) point inside the FLRW part varies as
\be \label{affine}
\overline{u}=c u+c_1
\ee
 where $c$ and $c_1$ are constants.
As a result $\langle \overline{x}^\pm| T_\eta^\chi  |\overline{x}^\pm \rangle =  0$, which says no quantum particle creation occurs from the IMOTS in 2-dimensional OS black hole. Any other collapsing models which keeps the linear function (\ref{affine}) between the two affine parameters also gives a zero term for quantum particle creation. Note that we have ignored the reflection of the star surface. This result is compatible with the result at \cite{javad-ellis} where the authors show that by the tunneling method that the IMOTS does not have any radiation.


\subsection{Back scattering in dynamic black holes}

In this subsection, we will show that like in the case of the Schwarzschild stationary black hole, the 2-dimensional dynamic black hole and s-wave approximation for 4-dimensional dynamic black hole can describe the main features of black hole radiation.\\

We write the metric in the form $ds^2=-2 f(u,v) du~dv+R(u,v)^2 d\Omega^2$ where $u=x^-$ and $v=x^+$ only cover outside the horizon where the observer is.
Without giving the details, the Klein-Gordon equation for a general wave  equation becomes
\be
\partial_+\partial_- \varphi +\frac{[\partial_+ R ~\partial_- \varphi+\partial_- R~\partial_+ \varphi]}{R}+\frac{\ell(\ell+1)f}{2R^2}\varphi=0.
\ee
Defining the function $\varphi=\frac{\Phi}{R}$, we get this form for the general wave equation:
\be
\frac{\partial_+\partial_- \Phi}{R} -\frac{\partial_+ \partial_- R}{R^2}\Phi+\frac{\ell(\ell+1)f}{2R^3}\Phi=0.
\ee
We call $V_{\ell}=-\frac{\partial_+ \partial_- R}{R^2}+\frac{\ell(\ell+1)f}{2R^3}$ the effective potential in the general spherically symmetric spacetime, and the wave equation reduce to this simple form
\be
(\frac{\partial_+\partial_-}{R} + V_\ell) \Phi=0.
\ee
In the special s-wave case, $\ell=0$ and the Klein-Gordon equation is
\be
\partial_+(R^2\partial_- \varphi)+\partial_-(R^2\partial_+ \varphi)=0
\ee
which can be written:
\be
\partial_+\partial_- \varphi +\frac{[\partial_+ R ~\partial_- \varphi+\partial_- R~\partial_+ \varphi]}{R}=\partial_+\partial_- \varphi +\frac{[\theta_+ ~\partial_- \varphi + \theta_- ~\partial_+ \varphi]}{2}=0
\ee
where $\theta_{\pm}=2\frac{R_\pm}{R}$.\\

Defining the Misner-Sharp mass for this metric as $M=\frac{R}{2}(1+2\frac{\partial_+ R~\partial_- R}{f})$ \cite{Hayward:1994}, the apparent horizon for this metric is located in
\be
R=2M~~~\Rightarrow~~~\partial_+ R~\partial_- R=0
\ee
The future and past apparent horizon are distinguished by $\theta_{+}=0$ and $\theta_{-}=0$. In the case of gravitational collapsing models we consider the future apparent horizon. If we write the general spherically symmetric matter tensor in the double null coordinates $(u,v)$ as
\be
T_{\mu \nu}= pr^2d\Omega^2 + T_{++}du du + T_{+-}du dv + T_{--}dv dv
\ee
then Einstein's equation give
\be
R\partial_+ \partial_- R +f/2+\partial_+R \partial_- R=4\pi R^2 T_{+-}.
\ee
Using this result in the general wave equation we get
\be \label{wave-g}
\frac{\partial_+\partial_- \Phi}{R} -\left(\frac{4\pi R^2 T_{+-}-f/2-\partial_+R \partial_- R}{R^3}-\frac{\ell(\ell+1)f}{2R^3}\right)\Phi=0.
\ee

We can change the coordinates to $(U,V)$ where metric becomes
\be
ds^2=-2 F dU~dV+R^2 d\Omega^2
\ee
which like the Kruskal coordinates are regular at the horizon and cover all spacetime. The coordinate transformation between these two coordinates gives
\be
f=F (\frac{dU}{du} \frac{dV}{dv}+\frac{dU}{dv} \frac{dV}{du}).
\ee
Without loss of generality we take $U = p(u)$ and $V = q(v)$. Since the $(u,v)$ coordinates only cover the exterior of the horizon and $u$ goes to infinity at the future apparent horizon, we get
\be
\frac{dU}{du}\rightarrow0 ~~~\Rightarrow~~~f\rightarrow0.
\ee
An easier way to see this is that one can write the general spherically symmetric metric in the Kodama foliation \cite{visser-kodama}
\be
ds^2=-e^{\psi(t,r)}(1-\frac{2 M(t,r)}{R})dt^2+\frac{dr^2}{(1-\frac{2 M(t,r)}{R})}+R(t,r)^2 d\Omega^2
\ee
Here $M$ is the Misner-Sharp mass and these coordinates cover outside the horizon. This metric can be written 
\be
ds^2=-e^{\psi(t,r)}(1-\frac{2 M(t,r)}{R})^2 (dt^2+\frac{dr^2}{(1-\frac{2 M(t,r)}{R})^2})+R(t,r)^2 d\Omega^2.
\ee
Then we have
\be
f= I(t,r) e^{\psi(t,r)}(1-\frac{2 M(t,r)}{R})^2
~,~~du dv=\frac{1}{I(t,r)}(dt^2+\frac{dr^2}{(1-\frac{2 M(t,r)}{R})^2})
\ee where $I(t,r)$ is an integration factor to ensure that the second term is a perfect differential. One can easily see that $f$ is zero on the horizon.
 Therefore, even in the general spherical symmetric model we can neglect the backscattering, $\ell\neq 0$ term in eqn.(\ref{wave-g}), and only consider the s-waves term near the future apparent horizon, given by $\theta_{+}=2\frac{R_+}{R}=0$. \\
 
 Unlike the Schwarzschild black hole where the general wave equation near the apparent horizon reduces to the 2-dimensional wave equation and only the first term in equation (\ref{wave-g}) takes  part, generally non-zero matter terms cause that the 2-dimensional wave equation can be different from the 4-dimensional wave equation near the future apparent horizon for generally spherically symmetric models. As a result of the above calculation, in the case that we have a zero matter term, we can neglect the effective potential term, the 4-dimensional general wave equation reduces to the 2-dimensional wave equation, and the general features of the radiation like its thermal character and temperature are the same. On the other hand we have discussed above that only isolated black hole (black holes which already have no flux or matter on the horizon) can radiate. Therefore, we can infer that the backscattering term can be neglected for all radiating black holes, and s-waves and the 2-dimensional wave equation can well describe the general features of the radiating black hole near horizon even in the non-static case.

\section{Conclusion and discussion }

We have shown that energy conservation,  required for virtual pair creation to be real, gives a strict limitation on which black holes can be radiative.  It was shown that general black holes that have a \emph{dynamical horizon} \cite{ashtekar-02} cannot have pair creation on the horizon, because they violate energy conservation. Only isolated horizons and quasi-isolated horizons (slowly evolving horizon) can create real pairs which are not 
forbidden.\\

Next, we have considered how the expectation value for the quantum stress tensor carries information about quantum effects in curved space time. Beside the zero point energy there are two important quantum effects, namely particle creation and vacuum polarization, which can be studied through the expectation value of the quantum stress tensor in curved space time. Generally there is no way to distinguish these two effects in curved space time,  but having a horizon and choosing suitable coordinates help us to identify these two effects in the expectation value of the quantum stress tensor. Existence of an event horizon so that the observer cannot see part of spacetime is one  criterion for  having particle creation in a space time. To identify particle creation in the quantum stress tensor, one must choose a vacuum state that covers both inside and outside the horizon; this is a mixed state for the observer.\\

In this way, we have studied particle creation and vacuum polarization in cosmological models. We have defined three different vacuum states for the de Sitter spacetime which are like the Boulware, Hawking-Hartle, and Unruh vacua in the black hole case. It was discussed that the Boulware like vacuum state describes the vacuum polarization in the expectation value of the energy momentum tensor. The Hawking-Hartle vacuum state is the state of a thermal equilibrium universe with the temperature $T=\frac{\hbar\kappa_c}{2\pi}$. The calculation of  the particle creation term by  the expectation value of the quantum stress tensor gives the same thermal radiation that Gibbons and Hawking have calculated \cite{Hawking-77} for the de Sitter stationary spacetime. It was shown that this thermal temperature is compatible with the temperature that is calculated by the affine null parameter approach. In addition, the different components of the vacuum polarization tensor are calculated in this frame.\\

 Different aspects of particle creation in the cosmological de Sitter were discussed. Since we only able to see the events which are inside our past light cone, the observable quantities must be written for that region. For example in the de Sitter static spacetime we have to calculate the quantities for a coordinate that just covers inside the horizon (like cosmological tortoise coordinate). For the cosmological de Sitter model, the Unruh like vacuum that we proposed is interesting for cosmological application because, like the Bunch-Davies vacuum, it covers also the eternally expanding cosmological part (upper triangle) of the de Sitter hyperboloid and gives quantities which an observer can measure. \\

For the black hole case, the global nature of vacuum states is discussed and it was shown that it is not possible to study the vacuum polarization and particle creation locally by using the expectation value of the quantum stress tensor. We have shown that particle flux from the quantum stress tensor gives the same temperature that the affine null parameter approach \cite{vanzo-09} gives for the asymptotically flat spacetime. We also investigated the nature of the particle term via Bogoliubov coefficients in the case that the event horizon has formed. The exponential relation between affine parameters in future and past null infinity implies thermal radiation properties for particles with positive energy and anti-particle with negative energy. This shows that the created particles are confined near the event horizon. We have shown that the vacuum polarization flux for comoving observer (relative to the $t=\frac{u+v}{2}$ coordinate) in the case of a spherically symmetric static compact star or black hole and de Sitter space time is zero.\\

As an example, using the expectation value for quantum stress tensor, we have shown that the particle creation surface cannot be attributed to the IMOTS surface in Oppenheimer-Snyder collapse.
Finally, it was shown that we can neglect backscattering, $\ell\neq 0$, terms and only consider the s-waves term near the future apparent horizon;  the s-wave term gives the important part of Hawking radiation in general spherically symmetric black holes.\\

This paper introduces ways for choosing good coordinates for separating particle creation from vacuum polarization in the case of general non-stationary metrics.  This will be of practical use in examining the black hole back-reaction problem \cite{back-reaction}.\\

\appendix
In the Appendices we  summarise the nice presentation of quantum stress tensors given by  Saida, Harada and Maeda \cite{saida07} that is used in what is presented in previous sections. Appendix A gives the 2-dimensional case and Appendix B  the 2-dimensional case.

\section{Two dimensional case}\label{A}
It has already been recognized for a few decades that many different methods of renormalization give equivalent results (see for example, chapters 6 and 7 in~\cite{birrel-book}).
We consider a minimally coupled massless scalar field $\phi$, whose stress-energy tensor is given by
\begin{equation}
T_{\mu \nu} =
\phi_{,\mu} \phi_{,\nu} - \frac{1}{2} g_{\mu \nu} \phi_{,\alpha} \phi^{,\alpha} \, .
\label{eq-app.two-c.stress}
\end{equation}
The background spacetime is described in double null coordinates $(u,v)$ as
\begin{equation}
ds^2 = -  D(u,v) \, du\, dv \, .
\end{equation}

The field $\phi$ satisfies the Klein-Gordon equation $\Box\phi = 0$.
When a coordinate system (not necessarily null) is specified to describe the differential operator $\Box$, we can find a complete orthonormal set $\{f_{\omega}\}$ for arbitrary solutions of $\Box\phi = 0$, where $\omega$ denotes the frequency of the mode function.
The positive frequency mode is the mode function $f_{\omega}$ which is constructed to satisfy the conditions, $\omega > 0$~, $\left( f_{\omega} , f_{\omega'} \right) = \delta(\omega - \omega')$~, $(f_{\omega} , f_{\omega'}^{\ast}) = 0$ and $\left( f_{\omega}^{\ast} , f_{\omega'}^{\ast} \right) = - \delta(\omega - \omega')$, where $(f,g)$ is the inner-product defined from the Noether charge of time translation of $\phi$ and $f_{\omega}^{\ast}$ is complex conjugate to $f_{\omega}$, called the negative frequency mode.\\

In two dimensional spacetimes, the positive frequency modes can be decomposed with respect to the direction of propagation.
In the double null coordinates, they are $f_{\omega}(u) = \exp(-i\,\omega\,u)/\sqrt{4 \pi \omega}$ and $f_{\omega}(v) = \exp(-i\,\omega\,v)/\sqrt{4 \pi \omega}$.
Then, the quantum operator $\phi$ is expanded by the complete orthonormal set of the positive and negative frequency modes as
\begin{eqnarray}
\phi(u,v) = \int_0^{\infty} d\omega\,
\left[\, a_{\omega}\, f_{\omega}(u)
+ a_{\omega}^{\dag}\, f_{\omega}^{\ast}(u)
+  b_{\omega}\, f_{\omega}(v)
+ b_{\omega}^{\dag}\, f_{\omega}^{\ast}(v) \,\right] \, .
\label{eq-app.two-expand}
\end{eqnarray}
The canonical quantisation presumes the simultaneous commutation relation between $\phi$ and its conjugate momentum, so that $\{a_{\omega}\}$ and $\{b_{\omega}\}$ are harmonic operators satisfying the commutation relations;
$[ a_{\omega} , a_{\omega'}^{\dag} ] = \delta(\omega - \omega')$ and $ [ b_{\omega} , b_{\omega'}^{\dag} ] = \delta(\omega - \omega')$ and all others vanish.
They define the Fock space of quantum states and give particle interpretation.
The vacuum state $\vac$ is defined as a quantum state satisfying $a_{\omega} \vac = b_{\omega} \vac = 0$ for all $\omega$.\\

If we choose different coordinates $(\bar{u},\bar{v})$, a natural orthonormal set of mode functions is $\{\bar{f}_{\omega}\}$, where $\bar{f}_{\omega}(\bar{u}) = \exp(-i\,\omega\,\bar{u})/\sqrt{4 \pi \omega}$ and $\bar{f}_{\omega}(\bar{v}) = \exp(-i\,\omega\,\bar{v})/\sqrt{4 \pi \omega}$.
Then the expansion \eqref{eq-app.two-expand} gives different harmonic operators $\{\bar{a}_{\omega}\}$ and $\{\bar{b}_{\omega}\}$. These harmonic operators define another vacuum state $\bvac \,(\,\neq \vac \,)$ if there arises a mixing of positive and negative frequency modes $\left( f_{\omega} , \bar{f}_{\omega'}^{\ast} \right) \not\equiv 0$ between the two coordinate systems.
Thus, even if a quantum state is initially set to be a vacuum state, this does not remain vacuum but corresponds to an excited state associated with the coordinate system natural to an observer at the final time if the mixing of positive and negative modes arises.
This will be interpreted as quantum particle creation in curved spacetimes.\\

The quantum expectation value of the stress-energy tensor $\bvacbra T_{\mu \nu} \bvac$ is
calculated by substituting the quantum operator \eqref{eq-app.two-expand} (after replacing $a_{\omega}$ and $b_{\omega}$ with $\bar{a}_{\omega}$ and $\bar{b}_{\omega}$) into the stress-energy tensor \eqref{eq-app.two-c.stress}.
However, $\bvacbra T_{\mu \nu} \bvac$ diverges even for flat background cases.
Therefore, we need to renormalise the stress-energy tensor.
We do not go into the details of the regularization method but only quote the result~\S 6.4 in~\cite{birrel-book},
\begin{eqnarray}
\tmn{\bm \bn} = \theta_{\bm \bn} + \frac{\hbar{\cal R}}{48\, \pi}\, g_{\bm \bn} \, ,
\label{eq-app.two-stress.1}
\end{eqnarray}
where $\tmn{\bm \bn}$ is the renormalized expectation value of $\bvacbra T_{\bm \bn} \bvac$, ${\cal R}$ is the Ricci scalar of the background spacetime, and $\theta_{\bm \bn}$ is a symmetric tensor whose components in the coordinate system $(\bu,\bv)$ on which the vacuum $\bvac$ is defined is given by
\begin{subequations}
	\label{eq-app.two-stress.2}
	\begin{eqnarray}
	\theta_{\bu \bu} &\defeq&
	-\frac{\hbar}{24 \pi} \left[ \frac{3}{2} \left(\frac{D_{,\bu}}{D}\right)^2
	- \frac{D_{,\bu \bu}}{D} \right] , \\
	\theta_{\bv \bv} &\defeq&
	-\frac{\hbar}{24 \pi} \left[ \frac{3}{2} \left(\frac{D_{,\bv}}{D}\right)^2
	- \frac{D_{,\bv \bv}}{D} \right] , \\
	\theta_{\bu \bv} &=& \theta_{\bv \bu} \equiv 0 ,
	\end{eqnarray}
\end{subequations}
where $D(\bu,\bv) = - 2\, g_{\bu \bv}$.
The renormalized expectation value $\tmn{\mu \nu}$ of $\bvacbra T_{\mu \nu} \bvac$ in the other coordinates $(u,v)$ is calculated from the above components through the usual coordinate transformation for tensor components,
\begin{eqnarray}
\tmn{\mu \nu} =
\frac{\partial x^{\bm}}{\partial x^{\mu}}\,
\frac{\partial x^{\bn}}{\partial x^{\nu}}\, \tmn{\bm \bn} \, .
\end{eqnarray}

\section{Four dimensional case} \label{B}

The renormalized vacuum expectation value of the  stress-energy tensor $\tmn{\mu \nu}$ of matter field in four dimensions may also be calculated 
via the canonical quantization formalism as shown for the two dimensional case in the previous section.
However the path integral quantization formalism is more convenient to summarize $\tmn{\mu \nu}$ on a four dimensional conformal spacetime.\\

The effective action $W$ of a quantum matter field $\phi$ on a spacetime of metric $g_{\mu \nu}$ gives the vacuum expectation value of a quantum stress-energy tensor.
$W$ can be evaluated by the path integral method and the vacuum state $\vac$ is specified by the Green function of $\phi$ used in evaluating the path integral.
However the precise path integral form of $W$ is not important here.
$W$ is decomposed into two parts as $W = W_{\rm ren} + W_{\rm div}$, where $W_{\rm ren}$ is the renormalized part and $W_{\rm div}$ is the divergent part.
The functional differentiation of $W_{\rm ren}$ gives the renormalized vacuum expectation value $\tmn{\mu \nu}$,
\begin{equation}
\tmn{\mu \nu} = \frac{2}{\sqrt{-g}}\, \frac{\delta W_{\rm ren}}{\delta g^{\mu \nu}} \, .
\label{eq-app.four-stress.1}
\end{equation}

We consider the case where the metric $g_{\mu \nu}$ is conformal to the other one as
\begin{equation}
g_{\mu \nu} = \Omega^2\, \tg_{\mu \nu} \, ,
\label{eq-app.four-conf.metric}
\end{equation}
and the matter field $\phi$ is a conformally coupled massless scalar field satisfying $(\Box - {\cal R}/6) \phi = 0$. On the other hand, we get from the  definition of functional differentiation,
\begin{equation}
W_{\rm ren} - \tW_{\rm ren} =
\int \frac{\delta W_{\rm ren}}{\delta g^{\alpha \beta}}\,\delta g^{\alpha \beta}\, d^4x \, ,
\label{eq-app.four-action}
\end{equation}
where $\tW_{\rm ren}$ is the renormalized effective action obtained from $W_{\rm ren}$ on replacing $g_{\mu \nu}$ by $\tg_{\mu \nu}$.
Then considering functional differentiation only by the conformal transformation, $\delta g^{\mu \nu} = - 2 g^{\mu \nu}\,\Omega^{-1}\, \delta\Omega$, the effective action is expressed as
\begin{equation}
W_{\rm ren} =
\tW_{\rm ren} - \int g^{\alpha \beta}\tmn{\alpha \beta}\,
\frac{\delta\Omega}{\Omega}\, \sqrt{-g}\, d^4x \, .
\end{equation}
Substituting this into eq.~\eqref{eq-app.four-stress.1}, we get
\begin{equation}
\tmn{\mu \nu} =
\frac{1}{\Omega^{2}} \ttmn{\mu \nu}
- \frac{2}{\sqrt{-g}} \, \frac{\delta}{\delta g^{\mu \nu}}\,
\int g^{\alpha \beta}\tmn{\alpha \beta}\, \frac{\delta\Omega}{\Omega}\, \sqrt{-g}\, d^4x \, ,
\label{eq-app.four-stress.2}
\end{equation}
where $\widetilde{\delta}_{\mu}^{\nu} = \tg_{\mu \alpha} \tg^{\alpha \nu}$, $g_{\mu \sigma}\tg^{\sigma \alpha} = \Omega^2\, \widetilde{\delta}_{\mu}^{\alpha}$ and the general relation,
\begin{equation}
g^{\mu \alpha} \frac{\delta}{\delta g^{\alpha \nu}} =
\tg^{\mu \alpha} \frac{\delta}{\delta \tg^{\alpha \nu}} \, ,
\label{eq-app.four-dif.rel}
\end{equation}
is used to get the first term of the right-hand side of eq.~\eqref{eq-app.four-stress.2}.
The trace $g^{\alpha \beta}\tmn{\alpha \beta}$ is usually called the conformal anomaly or the trace anomaly, and it is well known that the divergent part $W_{\rm div}$ gives the conformal anomaly as (see \S 6.3 in~\cite{birrel-book} for example)
\begin{equation}
g^{\alpha \beta}\tmn{\alpha \beta} =
\frac{\Omega}{\sqrt{-g}}\, \frac{\delta W_{\rm div}}{\delta\Omega} \, .
\end{equation}
Hence substituting this expression of the conformal anomaly into eq.~\eqref{eq-app.four-stress.2} and using eq.~\eqref{eq-app.four-dif.rel} and eq.~\eqref{eq-app.four-action}, on replacing $W_{\rm ren}$ by $W_{\rm div}$ we obtain
\begin{equation}
\tmn{\mu \nu}
= \frac{1}{\Omega^2} \ttmn{\mu \nu}
- \frac{2}{\sqrt{-g}}\, \frac{\delta W_{\rm div}}{\delta g^{\mu \nu}}
+ \frac{2\,\Omega^2}{\sqrt{-g}}\, \frac{\delta \tW_{\rm div}}{\delta \tg^{\mu \nu}} \, .
\end{equation}

The divergent part $W_{\rm div}$ can be evaluated from the Green function of the matter field $\phi$. We do not present the details of the calculation of $W_{\rm div}$, but quote only the result for $\tmn{\mu \nu}$ for the conformally coupled massless scalar field $\phi$ on the spacetime of metric \eqref{eq-app.four-conf.metric} (see~ \S 6.2 and \S 6.3 in~\cite{birrel-book} for detail),
\begin{equation}
\tmn{\mu \nu}
= \frac{\hbar}{\Omega^2}\, \ttmn{\mu \nu}
- \frac{\hbar}{2880 \pi^2}\left( \frac{1}{6}\,X_{\mu \nu} - Y_{\mu \nu} \right)
+
\frac{\hbar}{2880 \pi^2\Omega^{2}}
\left( \frac{1}{6}\, \widetilde{X}_{\mu \nu} - \widetilde{Y}_{\mu \nu} \right) \, ,
\label{eq-app.four-stress}
\end{equation}
where
\begin{subequations}
	\label{eq-app.four-XY}
	\begin{eqnarray}
	X_{\mu \nu} &\defeq&
	2\, \nabla_{\mu} \nabla_{\nu} {\cal R} - 2\,g_{\mu \nu}\, \Box {\cal R}
	+ \frac{1}{2}\,{\cal R}^2\,g_{\mu \nu} - 2\,{\cal R}\, R_{\mu \nu} \, , \\
	Y_{\mu \nu} &\defeq&
	- R_{\mu}^{\alpha} R_{\alpha \nu} + \frac{2}{3}\, {\cal R}\, R_{\mu \nu}
	+ \frac{1}{2}\, R_{\alpha \beta} R^{\alpha \beta}\, g_{\mu \nu}
	- \frac{1}{4} {\cal R}^2\,g_{\mu \nu} \, ,
	\end{eqnarray}
\end{subequations}
and $R_{\mu \nu}$ and ${\cal R}$ are the Ricci tensor and scalar with respect to $g_{\mu \nu}$ respectively, and $\widetilde{X}_{\mu \nu}$ and $\widetilde{Y}_{\mu \nu}$ are defined similarly with respect to the metric $\tg_{\mu \nu}$.
Equation \eqref{eq-app.four-stress} is the generalization of eq.~(6.141) in~\cite{birrel-book} to the general conformal spacetimes of metric \eqref{eq-app.four-conf.metric}.\\


{\bf Acknowledgments:}\\

JF would like to thank Alessandro Fabbri, Mohammad Mehdi Sheikh-Jabbari and Hassan
Firouzjahi for useful discussions and comments. GE and JF  thank the NRF for financial support.\\

\end{document}